\documentclass{bmcart}

\usepackage[utf8]{inputenc} 
\usepackage{amssymb}
\usepackage{setspace}
\usepackage{booktabs}
\usepackage{soul,color} 
\usepackage{subfigure}
\usepackage{amsmath}
\usepackage{float}
\usepackage{graphicx}

\newcommand{\nop}[1]{}

\startlocaldefs
\endlocaldefs

\begin{document}

\begin{frontmatter}

\begin{fmbox}
\dochead{Research}


\title{Novel Modelling Strategies for High-frequency Stock Trading Data}


\author[
  addressref={aff1},                   
  corref={aff1},                       
  noteref={n1},                        
  email ={xuekui@uvic.ca}   
]{\inits{XZ}\fnm{Xuekui} \snm{Zhang}}
\author[
  addressref ={aff1,aff2},
  noteref ={n1},
  email={y664huan@uwaterloo.ca}
]{\inits{YH}\fnm{Yuying} \snm{Huang}}
\author[
  addressref={aff3},
  email={kexu@uvic.ca}
]{\inits{KX}\fnm{Ke} \snm{Xu}}
\author[
  addressref={aff4},
  email={li.xing@math.usask.ca}
]{\inits{LX}\fnm{Li} \snm{Xing}}


\address[id=aff1]{
  \orgname{Math\&Stat Department at University of Victoria}, 
  \city{Victoria},                              
  \cny{Canada}                                    
}

\address[id=aff2]{%
  \orgname{Stat\&Actuarial Science at University of Waterloo},
  \city{Waterloo},
  \cny{Canada}
}
\address[id=aff3]{%
  \orgname{Economics Department at University of Victoria},
  \city{Victoria},
  \cny{Canada}
}
\address[id=aff4]{%
  \orgname{Math\&Stat Department at University of Saskatchewan},
  \city{Saskatchewan},
  \cny{Canada}
}


\begin{artnotes}
\note[id=n1]{Equal contributor} 
\end{artnotes}

\end{fmbox}


\begin{abstractbox}

\begin{abstract} 
Full electronic automation in stock exchanges has recently become popular, generating high-frequency intraday data and motivating the development of near real-time price forecasting methods. Machine learning algorithms are widely applied to mid-price stock predictions. Processing raw data as inputs for prediction models (e.g., data thinning and feature engineering) can primarily affect the performance of the prediction methods. However, researchers rarely discuss this topic. This motivated us to propose three novel modelling strategies for processing raw data. We illustrate how our novel modelling strategies improve forecasting performance by analyzing high-frequency data of the Dow Jones 30 component stocks. In these experiments, our strategies often lead to statistically significant improvement in predictions. The three strategies improve the  F1 scores of the SVM models by 0.056, 0.087, and 0.016, respectively.
\end{abstract}




\end{abstractbox}
%

\end{frontmatter}




\section*{Introduction}
\label{sec:introduction}
High-frequency trading (HFT) arises from increased electronic automation in stock exchanges, which features the use of extraordinarily high-speed and sophisticated computer programs for generating, routing, and executing orders \cite{b1,b2}. Investment banks, hedge funds, and institutional investors design and implement algorithmic trading strategies to identify emerging stock price surges \cite{b3}. The increase in transaction efficiency increases the complexity of limit order book (LOB) data. Compared with stock trading before electronic automation, more quote data are generated in the LOB during the high-frequency trading process. Extracting useful information and modelling the complexity of massive LOB data for precise stock mid-price predictions are empirical big data challenges for traditional time-series methods. For instance, \cite{b50} suggests that classical machine learning methods actually surpass traditional models in precision for financial time-series predictions compared to the ARIMA and GARCH models. As computational resources, sophisticated datasets, and larger datasets continue to expand in the financial field, scholars and practitioners have developed increasingly elaborate methods for analyzing complex financial markets. In particular, machine learning has gained popularity in the finance industry because of its ability to capture nonlinearity, effectiveness, and strong predictive power. Innovative studies have demonstrated promising results for a variety of tasks. For example, machine learning and other advanced models have been employed for financial data mining \cite{li2021integrated}, financial market microstructure investigations \cite{ b7, huang2017nonlinear}, and stock price analysis \cite{b10, wen2019retail}.


Quantitative analyses of financial price predictions are important because more accurate predictions lead to higher profits from trading strategies \cite{b11,b12}. The quality of the prediction depends on two major factors: (1) the choice of statistical learning method used to train the prediction model and (2) the choice of input for machine learning methods, i.e., the extraction of information from large raw data, such as input variables (predictors) and subsets of training samples. The majority of the literature \cite{b44,b45,b46} focuses on enhancing prediction accuracy with advanced machine learning or deep learning models, which address the first factor discussed above. However, to the best of our knowledge, little attention has been paid to the second factor. This issue motivated us to study how to extract useful information from large amounts of raw data as inputs for machine learning methods. Next, we explain the importance of pre-processing raw data as input predictors, common practices to extract features from raw data, and the issues we want to address.

Although high-frequency data offer new opportunities to learn high-resolution information at the nanosecond level for financial analysis, it creates new challenges in acquiring and utilizing massive amounts of information. Given the vast amount of high-frequency data records, it is impossible to consider the entire dataset because it is too computationally expensive. Furthermore, close observations in high-frequency data are highly correlated \cite{b47,b48}, which violates the independence assumption of most machine-learning models. Hence, it is critical to properly process the raw data and convert them into meaningful inputs for machine learning models. To address this issue, the common practice in the literature to process high-frequency raw data is to apply the event-based protocol \cite{b38,b39} together with the sampling strategy, which randomly sub-samples raw data at fixed events. Both the event-based protocol and the sampling strategy are forms of data thinning. Such approaches substantially reduce the size of the dataset and weaken the correlation among observations. However, this widely used data thinning approach has three disadvantages. First, data thinning compromises the advantage of high resolution high-frequency trading data. While reducing the data density, data thinning (i.e., event protocol and subsampling process) discards the inherent information between fixed events.  Second, randomness in the data thinning procedure (e.g., different starting points of events and sampling strides) affects the models' robustness and reproducibility. Third, long-term trends in price history could provide useful information for prediction, but they are rarely used in current models. High-frequency data over a long historical period are difficult to handle by most models because this leads to numerous correlated predictor variables, which dilutes the impact of all predictors in the model and creates severe collinearity problems. Researchers tend to construct scalar variables based on data at or close to a specific timestamp without leveraging information over a long-time scope \cite{b12,b26,b39,b49}. In this study, we propose three novel modelling strategies that aim to address these disadvantages to alleviate the insufficient use of high-frequency data and improve mid-price prediction performance.

To overcome the first disadvantage, we devise Strategy I, which uses a collection of variables that summarize and recover useful information discarded during data thinning. In response to the second disadvantage, our second strategy proposes a stock price prediction framework called `Sampling+Ensemble’. This framework consists of two steps: the first step fits training models on many random subsets of original samples, and the second step integrates results from all models fitted in the first step and subsequently generates the final prediction through a voting scheme. This strategy combines the `Sampling' step to reduce between-sample correlation and computational load in each subset, and the `Ensemble' step to increase precision and robustness of predictions. This strategy is flexible, as users can choose from a wide range of machine learning models as their processors (base learners) to analyze each data subset generated in the `Sampling' step. In real data experiments, we used the support vector machine (SVM) and elastic net (ENet) models as the base learners to obtain benchmark results for performance comparison. Owing to the ENet model’s automatic feature selection property, we identified the importance of the predictor variables by ranking the total number of times each predictor was selected in the `Sampling’ step. Finally, our third novel strategy introduces a new feature to high-frequency stock price modelling, which emphasizes the importance of considering longer-term price trends. The implementation of the functional principal component analysis (FPCA) method \cite{b40} helps compress information in historical prices from a long and ordered list of correlated predictors into a few orthogonal predictors. We customize features that capture long-term price patterns over the past day, and examine whether they improve the prediction model.  

The proposed method can be applied to high-frequency trading algorithms to achieve improved forecasting performance and informational efficiency. We illustrate the performance improvements of our three novel strategies using high-frequency intraday data on Dow Jones 30 component stocks from the New York Stock Exchange (NYSE) Trade and Quote (TAQ) database. Following the problem set up in previous work \cite{b12}, we treat mid-price movement prediction as a three-class classification problem (i.e., up, down, or stationary mid-price states) for every next 5th event in each random subset of training data. We forecast the Dow Jones 30 stock prices using various machine learning methods with and without each of our novel strategies and evaluate the improvement in prediction performance using our novel strategies.  We used precision, recall, and F1 score as performance metrics, which are widely used in the machine learning community. To investigate the uncertainty of our comparison, we repeated our experiments 100 times on different random subsets of the original data and compared the performance metrics (e.g., F1 scores) using the non-parametric Wilcoxon signed-rank test. Evaluation results show that our second strategy (Sampling+Ensemble) is `consistently' helpful, significantly outperforming the original models without this strategy in all 30 stocks, with up to a $0.23$ increase in F1 scores. Our first strategy (recovering the discarded information by the data thinning process) is often helpful, significantly improving the prediction performance in 24 out of 30 stocks. Our third strategy (modelling long-term price trends by FPCA) can sometimes help significantly improve prediction performance in 8 out of 30 stocks.  Note that whether our first or third method helps depends on the characteristics of the data. If the last observation of event windows always carries most of the information of the window, recovering the loss during data thinning cannot help. If the long-term trends of a stock are unstable, modelling longer-term trends cannot be helpful. Finally, the ENet models provide us with the most frequently selected features for predicting each mid-price direction, which is novel knowledge for extending the existing features set for high-frequency mid-price prediction for further studies.



In the remainder of this study, we describe the setup of our research problem and propose novel strategies for data preprocessing. Then, we provide a brief introduction to two machine learning methodologies that are used to illustrate our novel strategies. We demonstrate how our novel strategies improve the prediction performance using the TAQ data analysis results. Finally, we conclude the study and discuss its limitations.

\section*{Problem Setup}\label{problem_description}
This section introduces the research questions and defines the notations and evaluation criteria of model performance.

In this study, our goal is to predict mid-price changes based on high-frequency LOB data. A limit order involves buying or selling security at a specific price or better. A buy limit order is an order to buy at a current or lower price, while a sell limit order sells security at no less than a specific price. The LOB accepts both buy and sell limit orders and matches buyers and sellers in the market. The highest bid price, denoted by $P^{bid}$, is the best bid price, whereas the lowest ask price, denoted by $P^{ask}$, is the best ask price. Their average price defines the so-called mid-price, namely $P^{mid}=(P^{bid}+P^{ask})/2$, whose movement is predicted. Every new limit order submission from either the buyer or seller creates and updates a new entry in the limit order book. More specifically, if the best bid price or best ask price is updated in the LOB, our mid-price will be updated accordingly, which we define as a trading event.  

\begin{figure}[!ht]
    \centering
    \includegraphics[width=10cm,height=5cm]{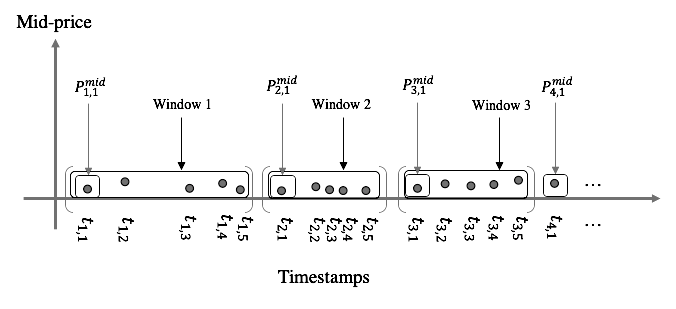}
    \caption{Illustration of event-based inflow framework with the length of each window k=5 events.}
    \label{Figure1}
\end{figure}

Assume that a dataset consists of chronologically recorded LOB events with an index ranging from $1$ to $N$. The occurrence of $N$ events (i.e., quotes) depends on the market. They do not have a steady inflow rate, i.e., the time intervals between two consecutive events vary tremendously from nanoseconds to minutes. Following the literature on event-based inflow protocols, we grouped every $k$ consecutive events in a window, which leads to $N/k$ windows for downstream analysis. Previous studies \cite{b38, b12} proposed various choices for the value of parameter $k$, ranging from 2 to 15. Such a value is not critical to illustrate the performance of our proposed novel strategies; therefore, we set $k=5$ for simplicity in our discussion. To handle the window-based data structure, as illustrated in Figure~\ref{Figure1}, we used a two-dimensional index system $(i,j)$ as the subscript for each event, where $i=1, \ldots, N/k$ denotes the $i$-th window and $j=1, \ldots, k$ denotes the event's position within the window. For example, the first LOB event in the $4$-th window occurred at time $t_{4,1}$ and had mid-price $P^{mid}_{4,1}$. To forecast this mid-price, we can use information from the previous windows.

The input data formats for supervised machine learning methods are significantly different from those for time-series methods. Time series methods consider data as a time-ordered vector of length $N$, whereas the input of machine learning methods consists of an $N$-dimensional outcome vector (same as a time series) and a predictor matrix of dimension $N \times p$. That is, each row of machine learning input data consists of an outcome (mid-price at a certain time-point), and $p$ predictors/features created from historical mid-price information before that time point (e.g., mid-prices of the last five trading events or two weeks of historical mid-prices traced back from the current time point).

Information on the relationship between consecutive observations of a time series is the most critical. Such information is equivalent to the outcome-predictor relationship within each row of machine learning input data. Hence, the order of the rows is not critical for machine learning methods. Moreover, the correlation between consecutive observations of time series methods is helpful for prediction, whereas correlations between rows in the machine learning data matrix void the independence assumptions of most machine learning methods. Therefore, the decorrelation among rows of the data matrix is important for machine learning methods. 

We defined three types of predictor variables to summarize the high-frequency historical information at different resolution levels under our proposed strategies. The first type consists of variables at the window level, which are fetched using one event (usually the last one) in each window as the standard classic features used in the literature. The details of this type of variable are presented in Table~\ref{f_tab_3} in the Data Cleaning and Multi-resolution Features Construction section. The second type consists of variables that capture micro-trends within each window and will be discussed in the section on our proposed Strategy I. The third type consists of variables that capture the trend of price change in long-term history and is discussed in the section on our proposed Strategy III.

Following \cite{b26}, we define the outcome variable based on the mid-price ratio between the average mid-price of all events in the current window $\sum_{j=1}^k P^{mid}_{i,j}/k$ and the last observed mid-price in its history $P^{mid}_{i-1,k}$. Using threshold values, we convert this ratio into a three-class categorical variable that represents three possible stock mid-price movement states: upwards, downwards, and stationary. Specifically, the outcome variable $Y_i$ of the $i$th record (or window) is defined as follows: 
\begin{equation}
Y_i=
 \begin{cases}
\text{Upwards,}& \text{if}~ \dfrac{\sum_{j=1}^k P^{mid}_{i,j}/k}{P^{mid}_{i-1,k}}>1+\alpha\\
\text{Downwards,}&\text{if}~ \dfrac{\sum_{j=1}^k P^{mid}_{i,j}/k}{P^{mid}_{i-1,k}}<1-\alpha\\
\text{Stationary,}& \text{otherwise}
\end{cases} \label{price_definition}  
\end{equation}
where $\alpha$ is the parameter that determines the significance of the mid-price movement. In practice, we suggest choosing the value for $\alpha$ using two rules. (1) The value should be large enough to make it meaningful in practice so that high-frequency trading decisions based on such $\alpha$ values can make a profit. (2) The value cannot be too large, so we have enough training data to model the ``upwards" and ``downwards" movements of stock prices.

Machine learning methods require a different data format than time-series methods do. The time series involves handling an $N$-dimensional vector indexed by time order. Time order is essential for time series forecasting because it predicts future values based on previously observed values. In contrast, machine learning methods predict future responses based on input features. In machine-learning methods, the temporal information contained in the time order of observations (of time-series methods) is converted into an outcome-predictor relationship within each row of the input predictor matrix (of machine-learning methods). In other words, the historical information is included in the predictor matrix with various resolutions, as stated above. Therefore, machine-learning methods do not require correlation information between consecutive observations. A strong correlation between the rows of the data matrix should be avoided to satisfy the independence requirements of most machine learning models. However, when the window size is not sufficiently large, the mid-price or converted categorical outcome $Y_i$ might be highly correlated with their adjacent records. We propose our first two strategies to address this issue, and discuss these strategies in the Novel Strategy section. 

To evaluate whether our novel strategies can improve prediction performance, for each fitted model, we calculated its recall, precision, and $F_1$ score, which are widely used performance metrics by the machine learning community. Our novel strategy is considered helpful if it leads to a positive change in performance metrics. The recall and precision metrics are defined as follows:
\begin{equation}
    Recall=\dfrac{TP}{TP+FN},~
    Precision=\dfrac{TP}{TP+FP}
\end{equation}
where TP is the number of true-positive predictions (e.g., correctly predict `upwards' as `upwards'), FN is the number of false-negative predictions (e.g.,  incorrectly predict `upwards' as `not upwards'), and FP is the number of false-positive predictions (e.g., incorrectly predict `not upwards' as `upwards'). Both recall and precision are performance metrics that are commonly used in classification tasks. Recall denotes the
proportion of true-positive cases that are correctly labelled as positive, while precision denotes the proportion of predicted-positive cases that are correctly labeled as true-positive. A good classification model performance aims to achieve a relatively high recall and precision simultaneously. Usually, when we analyze the results, we can either investigate and compare one measure when the other measure is at a fixed level or we can combine these two metrics into one. In this study, we used $F_1$ score, the harmonic average of precision and recall, as a single measurement of the classification task:
\begin{equation}
    F_1=\dfrac{2\times Recall \times Precision}{Recall + Precision}
\end{equation}

\clearpage

\section*{Novel Strategies}\label{strategy}
 In this section, we propose three novel strategies, describe the issues that they resolve, and explain the mechanisms behind them. Our objective is to preprocess high-frequency raw data into appropriate inputs for machine learning methods. All three novel strategies are independent of each other and can be applied separately or in combination. These strategies are not limited to mid-price prediction, but open avenues for high-frequency data applications in other fields. 
 
\subsection*{Strategy I: Recover Information in Data Thinning\label{strategyI}}

The aforementioned event-based inflow serves as a data-thinning strategy for high-resolution observations, which uses only one event (usually the last event) within each window. Using fewer events weakens the correlation between successive observations and reduces computational costs by shrinking the size of the dataset. However, each window carries much more useful information that can be captured by only one record. In particular, the records in the last window provide the most useful information for forecasting future prices. Using only one record in that window can result in significant loss of information. 

Our first strategy is to define a few new variables to recover discarded useful information within each window. Although observations within an event window can be highly correlated and carry redundant information, their trend can be helpful in predicting the movement of the next mid-price. Instead of using the features solely built by the ``record'' events, we included new variables to extract and generalize summary features within each window. More specifically, we proposed an extensive collection of input features based on information that can be extracted from events within each window, as depicted in Figure~\ref{Figure1}. The feature set contains features such as mean, variance, range mid-price observations, trade intensity, volatility, market depth, and bid-ask spread. As it summarizes the financial characteristics within each window, we call the new set of features as ``within-window high-frequency variables''. Detailed descriptions and calculation formulas of these variables are summarized in Table~\ref{t:windowVar}. 
\begin{table}[!ht]
\centering
\caption{Definition of Within-window variables with illustration of the feature extraction to predict the mid-price movement of the $i$-th window, $i=3,\dots, N/k$, where $P^{mid}_{i,\cdot}$ indicates the mid-price sequence within the $i$-th window, $(P^{mid}_{i,1},\dots, P^{mid}_{i,k})$. Each window contains k=5 events. \label{t:windowVar}}
\begin{tabular}{ll} 
\toprule
Definition                                                     & Description                        \\ 
\hline
$V_1=(P_{i-1,k}^{bid}-P_{i-1,1}^{bid})/P_{i-1,1}^{bid}$                     & best bid price difference return   \\
$V_2=(P_{i-1,k}^{ask}-P_{i-1,1}^{ask})/P_{i-1,1}^{ask}$                     & best ask price difference return   \\
$V_3=(P_{i-1,k}^{bid}-P_{i-1,1}^{ask})/P_{i-1,1}^{ask}$                     & bid-ask spread crossing return     \\
$V_4=\sum_{j=1}^{k}P_{i-1,j}^{ask}/k$                           & mean best ask price  \\
$V_5=\sum_{j=1}^{k}P_{i-1,j}^{bid}/k$                           & mean best bid price  \\
$V_6=\sum_{j=1}^{k}P^{mid}_{i-1,j}/k$                                  & mean mid-price       \\
$V_7=\sum_{j=1}^{k}V^{ask}_{i-1,j}$                                      & best ask price market depth        \\
$V_8=\sum_{j=1}^{k}V^{bid}_{i-1,j}$                                      & best bid price market depth        \\
$V_9=  \sqrt{Var(P^{mid}_{i-2,\cdot}, P^{mid}_{i-1,\cdot})}$                                         & within-window volatility of two previous windows        \\
$V_{10}=1/(t_{i-1,k}-t_{i-1,1})$                  & trade~intensity       \\
\toprule
\end{tabular}
\end{table}

This new collection of features can capture more temporal information and complement the variable set that is constructed based on the ``record'' observations. $V_1$ and $V_2$ are two types of returns, aiming to measure the percentage of price changes in the best bid price and the best ask price compared to the counterparts of their previous ``record'' event. $V_3$ denotes the bid-ask spread crossing return, which is an indicator of potential arbitrage profits. For example, a trader makes a profit when he buys the asset at time $t_{i-1,1}$ with the lowest ask price and sells it at time $t_{i-1,k}$ with the highest bid price. $V_4,~V_5$ and $V_6$ are the mean values of the best ask price, best bid price, and mid-price, respectively, among the five events within a window. The summed quantity quoted at the best bid and ask prices, revealing the market depth, is calculated in $V_7$ and $V_8$. The standard deviation of mid-price changes is also known as price volatility. In $V_9$, we measure within-window volatility by calculating the standard deviation among all events in the two previous windows. The utilization of events from two windows is preferable because the computed standard deviations of the events from only one window are most likely to be zero because of subtle volatility. The time length of the $(i-1)$-th window is determined by the time difference between the first and last events in that window, namely, $t_{i-1,k}-t_{i-1,1}$. This represents the actual trading time for five events to occur prior to the given ``record'' event in $i$th window. Therefore, its reciprocal, as computed in $V_{10}$, manifests the transaction intensity of the given window.

\subsection*{Strategy II: ``Sampling + Ensemble'' Model}\label{strategy2}
The characteristics of high-frequency trading data lie in their massive trading volume and high data dependence among observations. Although it provides us with high-resolution data to train our models, high-frequency data also lead to challenges in analyzing such data. The massive amount of data is not manageable by most modern computers, and the high correlation among adjacent observations violates the independence assumptions of most machine-learning models. To address these challenges, the current standard approach is data thinning by randomly sampling event windows. However, such a sampling approach leads to further information loss and reduces the reliability of the results (i.e., depending on the random subset selected for data analysis). To improve robustness and address information loss, we propose a second strategy that combines the sampling approach with ensemble machine learning. Specifically, we used the bagging approach (a popular ensemble machine learning method) to combine many models fitted on various random subsets of the original training data. 


We propose our second modelling strategy, ``Sampling+Ensemble”, which retains the benefits of the sampling approach discussed above and addresses its robustness and information loss issue. Specifically, we randomly generated 100 subsets of training data, fitted a prediction model on each data subset, and used the average prediction of all 100 models as our final output. Each training subset uses only a portion of the original data, but the union of 100 subsets can cover the majority of the original data to avoid information loss. Integrating prediction results from models fitted on various subsets of original data can average the impact of subset selection, provide more robust results, and utilize more information/data than a model fitted on a single subset.

Note, using `100' random subsets is our empirical choice after testing it on many data sets. Using too many subsets can substantially slow down the analysis but results in little improvement in prediction performance. Using too few subsets cannot achieve the desired robustness, and more information is lost in the data. Users can adjust this setting according to their specific problems, if needed.

\subsection*{Strategy III: Combination of Long-term and Short-term Resolution \label{strategy3}}
The essential information used for predicting future mid-price movements is historical observations of mid-prices. In most prediction models, modelling a longer history of mid-prices requires including more historical observations as model predictors. Most machine learning models only fit a limited number of predictors, and hence cannot model how long-term history affects future mid-prices. This disadvantage is worse when analyzing high-frequency trading data because higher frequency data leads to redundant events observed over a long period. Thus, most current prediction models for high-frequency trading data in the literature utilize only information from a short-term history. This motivated us to propose a third strategy that considers long-term price effect features to enhance information capacity in the current feature set. 

Strategy III uses Functional Principal Component Analysis (FPCA) to reduce the dimensions of long-term history data before including them in the prediction model. FPCA is a dimension-reduction method similar to the Principal Component Analysis (PCA) method. PCA considers observations as vectors whose order is interchangeable, whereas FPCA handles observations as functions with interchangeable time orders. In other words, FPCA can utilize information from the mid-price sequence. We chose FPCA instead of PCA because the temporal information in the mid-price history plays a critical role in its prediction. In the prediction model, we represented the trends in the long-term history using a few FPCA scores instead of a long list of predictors (raw observed mid-prices).

In this study, we consider the long-term price effect of one-day history based on our empirical results and calculate the top Functional Principal Components (FPC) scores that account for $99.9\%$ of information in these historical trends. Users can use historical data of customized durations (e.g., 3-days or one week) according to their research objectives. Note that we expect that the long-term impact variables will uplift the prediction performance if the trajectory of the mid-price movement has low volatility. By contrast, if the mid-price movement trajectory is unstable and has rapid reversal or momentum, incorporating long-term impact variables in the prediction model will backfire. We could include both long- and short-term variables in the preliminary model and use machine learning methods to decide whether to retain the long-term variable in the final model. For example, the elastic net model has feature selection functionality and is suitable for this type of task. Users can also decide whether to include long-term variables manually, according to the stocks' recent qualitative characteristics.

A detailed description of FPCA can be found in \cite{b40}, \cite{b41} and \cite{b42}. In this study, we briefly introduce the key concept of FPCA. The FPCA projects the input trajectories of mid-price history to the functional space spanned by orthogonal FPC, and functional scores are the corresponding coordinates in the transformed functional space. Each component of the functional score vector is related exclusively to one FPC. The first FPC accounts for the largest proportion of variance in the data. By analogy, the next FPC explains the proportion of the rest of the variance after excluding previously generated FPCs. Based on our empirical findings, the first few FPCs account for most of the variance in the one-day historical data. Therefore, we reduce the dimensionality of data by choosing a few top FPCs that explain the majority of variance in the data and use the corresponding FPC scores to replace raw data with a long history of mid-prices. We denote $s_{ij}~ (i=1,\dots, N; j=1,\dots,K)$ the $j^{th}$ FPC score of the $i^{th}$ trajectory in the data, which is defined by
\begin{equation}
    s_{ij}=\int \delta_j (t) \big [X_i(t)-\bar{X}(t)\big ] dt
\end{equation}
subjected to constraints
\begin{equation}\label{constraint}
 \int \delta_j (t) \delta_h (t) dt=
\begin{cases}
1, j=h\\ 0,j \neq h
\end{cases}   
\end{equation}
where t is the continuous timestamp, $X_i(t)$ is the mid-price of the $i^{th}$ trajectory at time $t$, and $\bar{X}(t)=\sum_{i=1}^N X_i(t)/N$ is the point-wise mean trajectory of all samples in the data. The first principal component as the weight function is specified by $\delta_1 (t)$, which maximizes the variance of the functional scores $s_{i1}$ subject to Eq~(\ref{constraint}). The second-, third-, and higher-order principal components $\delta_j (t)$ are defined in the same way, but each of them explains the variance of the data in addition to the previously established ones, and they also need to meet the same constraint that requires all the functional principal components to be orthogonal.

\clearpage

\subsection*{Using proposed strategies with Machine Learning Methods \label{method}} 
Our proposed novel strategies focus on preprocessing raw HFT data into input data for machine-learning methods. In this study, we used two machine-learning methods to illustrate the application of our strategies. In this study, we illustrate the application of our strategies using the two most popular machine learning methods: Supporting Vector Machines (SVM) \cite{b17, b18, b19} and Elastic Net (ENet) \cite{b24}. 

The SVM model categorizes the response variables into two classes according to their input features. To achieve this goal, the SVM maps training samples to space and constructs a hyperplane along with two supporting vectors based on the training data. SVM further separates samples from the two classes using the hyperplane by maximizing both margins between the two supporting vectors. The new data points were then mapped into the same space and classified according to their position on the hyperplane. The ENet is a regularized linear regression model. It has a LASSO penalty and a Ridge penalty on the regression coefficients. The LASSO penalty can force irrelevant predictor coefficients to zero, thereby achieving automated feature selection. The Ridge penalty can shrink all predictor coefficients towards zero, which helps to address the collinearity and overfitting problems. The ENet model has two parameters that control the strength of two penalties: $\lambda$ controls the overall strength, and $\alpha^*$ controls the weights between two penalties. In the Empirical Application section, we provide details on how to select the values of these two parameters. In Section 1 of the supplementary document, we provide a more detailed description of the two methods. 

Next, we used real data to show how our novel strategies improve the prediction performance of these two machine learning models.


\section*{Empirical Application}


\subsection*{Data} \label{data}

To illustrate the prediction performance improved by each of our novel strategy, we acquired data from the New York Stock Exchange (NYSE) Daily Trade and Quote database (TAQ), which consists of high-frequency intra-day quote and trade data for NYSE-traded securities in all public exchanges nationwide. The intraday order-level data comprise the continuous trading time between 9:30 am to 4:00 pm every trading day from June to August 2017 (64 trading days), with nanosecond (one billionth of a second) timestamps (e.g., HHMMSSxxxxxxxxx). We focused on the component stocks of Dow Jones 30\footnote{WBA replaced GE on June 28, 2018; DOW replaced DWDP, on March 27, 2019}. The Dow Jones 30 includes the most prominent publicly traded companies in the U.S., representing a strong assessment of the market's overall health and tendencies. The details of these 30 stocks are listed in Appendix Table~\ref{d_tab_1}, which consists of different industry sectors such as conglomerates, financial services, and information technology.

\begin{figure}[!htbp]
    \centering
    \includegraphics[width=12cm,height=7cm]{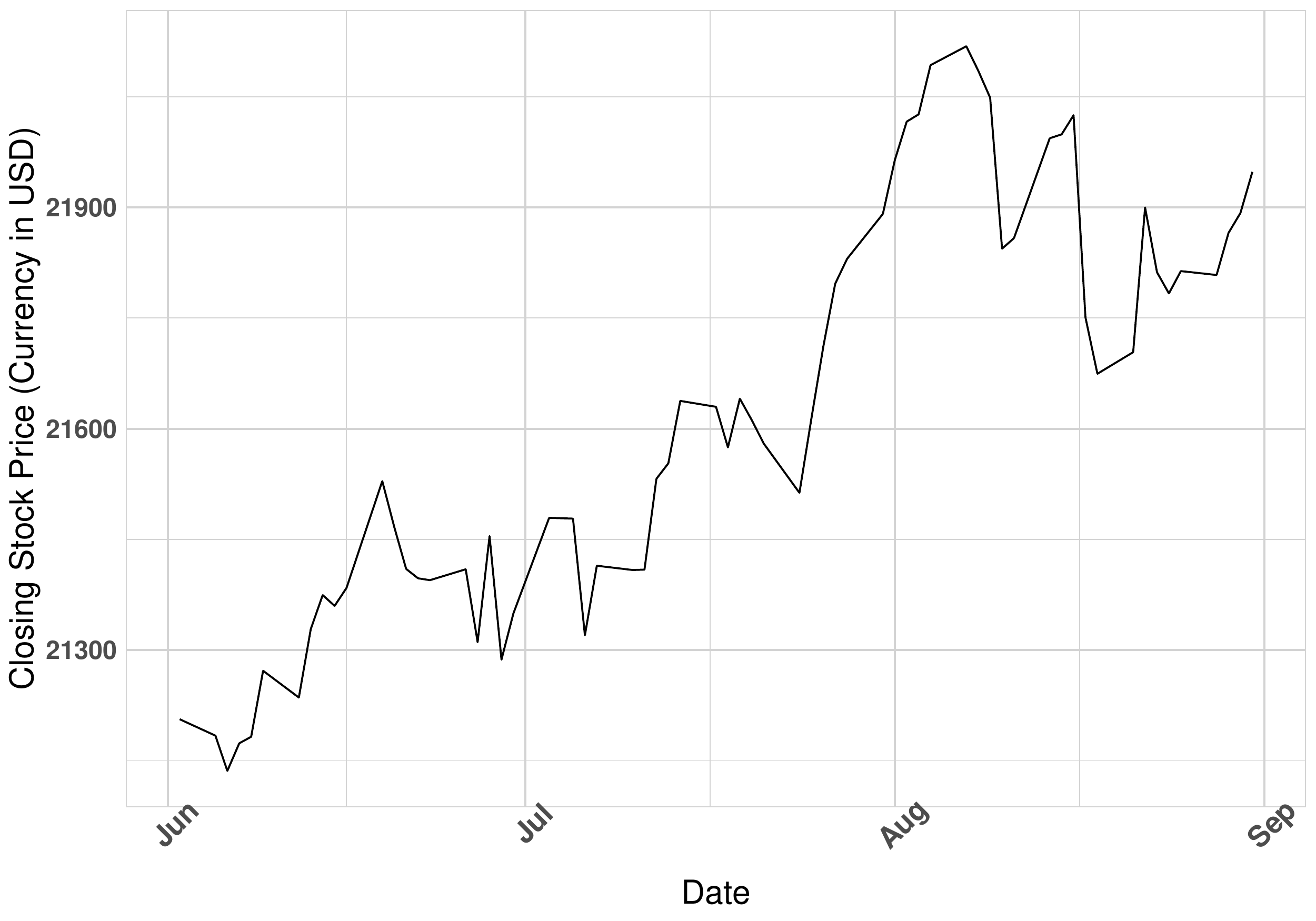}
    \caption{2017 daily adjusted closing stock price of Dow Jones 30}  \label{dj30}
\end{figure}

We plot the daily adjusted closing stock price of the Dow Jones 30 index during our sample period in Figure~\ref{dj30}. The Dow Jones 30 index increased by 3.42\% during the three-month study period. There were no extreme price movements in the Dow Jones 30 index during our sample period. In Table~\ref{summary_stat}, we present the summary statistics of market capitalization, trading volume, bid-ask spread, mid-price, and market depth. The average market capitalization of the Dow Jones 30 stock at the beginning of our sample period was USD 368 billion. The Dow Jones 30 stocks are highly liquid with an average bid-ask spread of 1.637 basis points and an average market depth of 2950 shares.

\begin{table}[ht]
\caption{Summary statistics for the full sample period}\label{summary_stat}
\centering
\begin{tabular}{rrrrrr}
\toprule[1pt]
 & MktCap  & Volume & Spread & Midprice  & Depth  \\
 & (\$billion) &  (million) &  (bps) & (\$ per share) &(shares) \\
  \hline
Mean & 367.878 & 9.801 & 1.637 & 98.296 & 2950.071 \\
  Median & 232.265 & 6.430 & 1.253 & 87.985 & 1200.000 \\
  Std Dev & 534.232 & 9.188 & 4.322 & 45.439 & 5795.606 \\
  Min & 37.162 & 0.955 & 0.000 & 30.320 & 200.000 \\
  Max & 2405.000 & 75.611 & 2196.646 & 246.415 & 580200.000 \\
\toprule[1pt]
\end{tabular}
\end{table}

We turn the response variable (i.e., stock mid-price) into a three-class categorical variable for the prediction. We used a small value $\alpha=10^{-5}$ in Eq~(\ref{price_definition}) to ensure that the stationary state has a similar sample size to the two other states and to make the upward or downward movements noticeable changes in stocks' mid-prices. The $\alpha$ value depends on stock volatility. We also experimented with two other values, $10^{-6}$ and $10^{-4}$. We found that a value around the $10^{-5}$ threshold is suitable for most of our categorical responses to obtain a balance. The value $\alpha=10^{-4}$ leads to an extreme imbalance in most stocks, whereas $\alpha= 10^{-6}$ leads to similar imbalance as our choice of $\alpha=10^{-5}$, but it is less financially significant. For more details on the proportion of response Y based on $\alpha$, please refer to Tables S1–S4 in Section 2 of the supplementary document. Moreover, we used a stratified sampling approach to construct our training datasets and kept the ratio of labels for mid-price (i.e., upward, downwards, and stationary) 1:1:1 in each training subset of data. This manipulation approach improves the data balance and makes it easier to compare the prediction performance. 

 \clearpage

\subsection*{Data Cleaning and Multi-resolution Features Construction}

Following \cite{b27}, we cleaned our data to ensure legitimacy and consistency using four steps. (i) Eliminate records beyond the exchange opening time from 9:30 am to 4 pm; (ii) Eliminate quotes with negative price, size, or bid price greater than the ask price; (iii) Eliminate trades with zero quantities; (iv) Eliminate trades with prices more than (less than) 150$\%$ (50$\%$) of the previous trade price; and exclude quotes when the quoted spread exceeds 25$\%$ of the quote midpoint or when the ask price exceeds 150$\%$ of the bid price.
 
Next, we standardize the variables using winsorization and normalization. The main purpose of winsorization and normalization is to remove extreme values and alleviate the impacts of different scales or units of the predictors. In the winsorization step, we removed extreme values detected by the same approach as used in the box plot method. We first computed the first and third quantiles (Q1 and Q3) of our training sample and calculated the interquartile range (IQR) equals to Q3-Q1. Next, we replaced the observations falling outside [Q1-1.5IQR, Q3+1.5IQR] with the lower bound Q1-1.5IQR and upper bound Q3+1.5IQR, respectively. The normalization step standardizes each variable using its mean values and standard deviations from the training samples. 

From the cleaned HFT data, we constructed features at three different resolutions for the prediction models based on machine learning methods: (1) window-level features used by standard methods in the literature, (2) within-window features proposed in Strategy I listed in Table~\ref{t:windowVar}, and (3) long-term history represented by FPCA scores as proposed in Strategy III.

The window-level variables are presented in Table~\ref{f_tab_3}. Variables $V_{11}$ to $V_{15}$ are the best bid price/volume, best ask price/volume, and mid-price, respectively, which are fetched directly from the LOB data. These are classic economic variables that measure changes in commonly used financial indicators before the ``record'' event. $V_{16}$ is an indicator of the bid-ask spread return. The bid-ask spread refers to the difference between the best ask price and the best bid price at the same timestamp. Typically, a narrow bid-ask spread exhibits a high volume of demand. On the contrary, a wide bid-ask spread may imply a low demand; therefore, it has an impact on the discrepancy in the asset price. Moreover, we measure the stock spike in features $V_{17}$ to $V_{21}$ through the average time derivatives of price and volume computed over the most recent second \cite{b12}. This helps us track whether there are relatively large upward or downward changes in trading prices and volumes within a very short period of time. Similarly, we measured the short-term average arrival rate by counting the number of quotes from both sides during the most recent second in feature $V_{22}$. 

Note that compared with variables used in other popular methods, such as \cite{b12}, we did not include window-level variables that require depth levels larger than 1. Because our LOB records from the NYSE dataset only provide information about the best bid and ask, i.e., depth$=1$, these variables cannot be derived from our data.

\begin{table}[h!]
\centering
\caption{Definition of standard variables with illustration of the feature extraction to predict the mid-price movement of the $i$-th window, $i=3,\dots N/k$. Each window contains k=5 events.\label{f_tab_3}}
\begin{tabular}{ll} 
\toprule
Definition                                                     & Description                        \\ 
\hline
$V_{11}=P^{ask}_{i-1,k}$                                 &best ask price       \\
$V_{12}=P^{bid}_{i-1,k}$                                                                                &  best bid price                                  \\
$V_{13}=P^{mid}_{i-1,k}$                                                                               &   mid-price                                 \\
$V_{14}=V^{ask}_{i-1,k}$                                                                                 &      best ask volume                              \\
$V_{15}=V^{bid}_{i-1,k}$                                                                              &    best bid volume                                \\
$V_{16}=(P_{i-1,k}^{ask}-P_{i-1,k}^{bid})/P^{mid}_{i-1,k}$                      & bid-ask spread return              \\
$V_{17}=dP^{ask}/dt$           & best ask price derivative       \\
$V_{18}=dP^{bid}/dt$                                                                          & best bid price derivative                                   \\
$V_{19}=dP^{mid}/dt$                                                                          &      mid-price derivative                               \\
$V_{20}=dV^{ask}/dt$                                                                           &    best ask volume derivative                                \\
$V_{21}=dV^{bid}/dt$                                                                           &  best bid volume derivative                                  \\
$V_{22}=$$\#$ \text{of events in the last second}        & arrival rate         \\
\toprule
\end{tabular}
\end{table}



\clearpage

\subsection*{Design of Benchmark Study Using Real Data}\label{result}
We conducted a benchmark study to evaluate the prediction performance of each of the proposed strategies. This study uses all component stocks of Dow Jones 30 from our NYSE data. From each stock, we randomly sampled 8000 records as training sets to train the prediction model and 2000 records as the testing set to evaluate the prediction performance. The evaluation results can be severely affected by sampling bias, i.e., the records were randomly selected in this experiment. To remove unwanted selection bias, we repeated this experiment 100 times by drawing 100 different random training and testing sets. We conclude based on 100 experiments, by averaging the sampling bias and learning the uncertainty in our evaluation. 

In the training sets, we fitted four types of models to investigate the prediction performance improvement made by each of the predicted strategies. First, we fit an SVM model using all predictor variables at three different resolutions, including the standard window-level feature set (shown in Table~\ref{f_tab_3}), the ``within-window'' feature set (shown in Table~\ref{t:windowVar}), and the FPCA scores discussed above. This model utilizes Strategies I and III. We considered this model as a baseline and compared it with the other three models. Next, we fit two reduced SVM models by removing the ``within-window'' features (Strategy I) and FPCA scores (Strategy III) from the baseline model, respectively. Comparing these two reduced models to the baseline model, we can evaluate the change in prediction performance caused by Strategies I and III. Finally, we utilized 100 baseline models on different random subsets of data to construct an ensemble model and compared it with the baseline model to evaluate the usefulness of Strategy II. In summary, our experiment consists of four types of models. They are the baseline model (Strategies I, III), the ensemble model (Strategies I, II, III), the``within-window'' model (Strategy I), and the FPCA model (Strategy III).

In the testing sets, we applied the trained SVM models to predict the mid-price movement of each record using historical trade data. Then, we compared the predicted movement with the observed movement to calculate the prediction performance criteria: recall, precision, and F1 score. We used the F1 score as the major criterion. To evaluate the performance improvement of the proposed strategy on each testing dataset, we calculated the F1 score difference of the two corresponding models (with and without that strategy). For example, the performance of Strategy I can be evaluated by the F1 difference between the baseline model (Strategies I and III) minus the FPCA model (Strategy III). In total, this leads to 90,000 F1 score differences obtained from combinations of the 3 strategies, 30 stocks, and 100 experiments. Furthermore, to learn the performance of our strategy with other machine-learning methods, we repeated these experiments with a different learner, ENet models, and all SVM models were replaced.

When training the SVM models, the kernel function used in this study was the polynomial kernel $\kappa (x_i,x_j)=(x_i \cdot x_j +1)^d$ with d = 2 and the constraint parameter $C = 0.25$ as suggested in \cite{b12}. When training the ENet model, we chose the values of parameters $\lambda$ and $\alpha^*$ in Eq. ~(8) of the supplementary document using a two-layer cross-validation (CV) approach. We applied a 5-folds CV grid search to each training sample. The regularization parameter $\lambda$ is evenly spaced on the log-scale range of $10^{-8}$ and 5 at 100 values, meanwhile, with a fixed $\lambda$, we searched for $\alpha^*$ values from a sequence of 4 values ranging from 0.2 to 0.8 with a stride of 0.2. We evaluated each combination of the two parameters and then determined the $\lambda$ and $\alpha^*$ that yield the best model performance \cite{b43}.

The evaluation results are presented in the following subsections and Appendix. In addition to the prediction performance, we also evaluated the importance of each handcrafted feature for the mid-price movement prediction task according to its frequency of being selected by the ENet model, the details of which are provided in Figure~\ref{Figure6}. 

\subsection*{Performance Evaluation of Proposed Strategies \label{performance_evaluation}}


\begin{figure}[!htpb]
\centering
\subfigure[Results with utlization of the SVM model\label{Figure5}]{
\includegraphics[width=120mm]{ 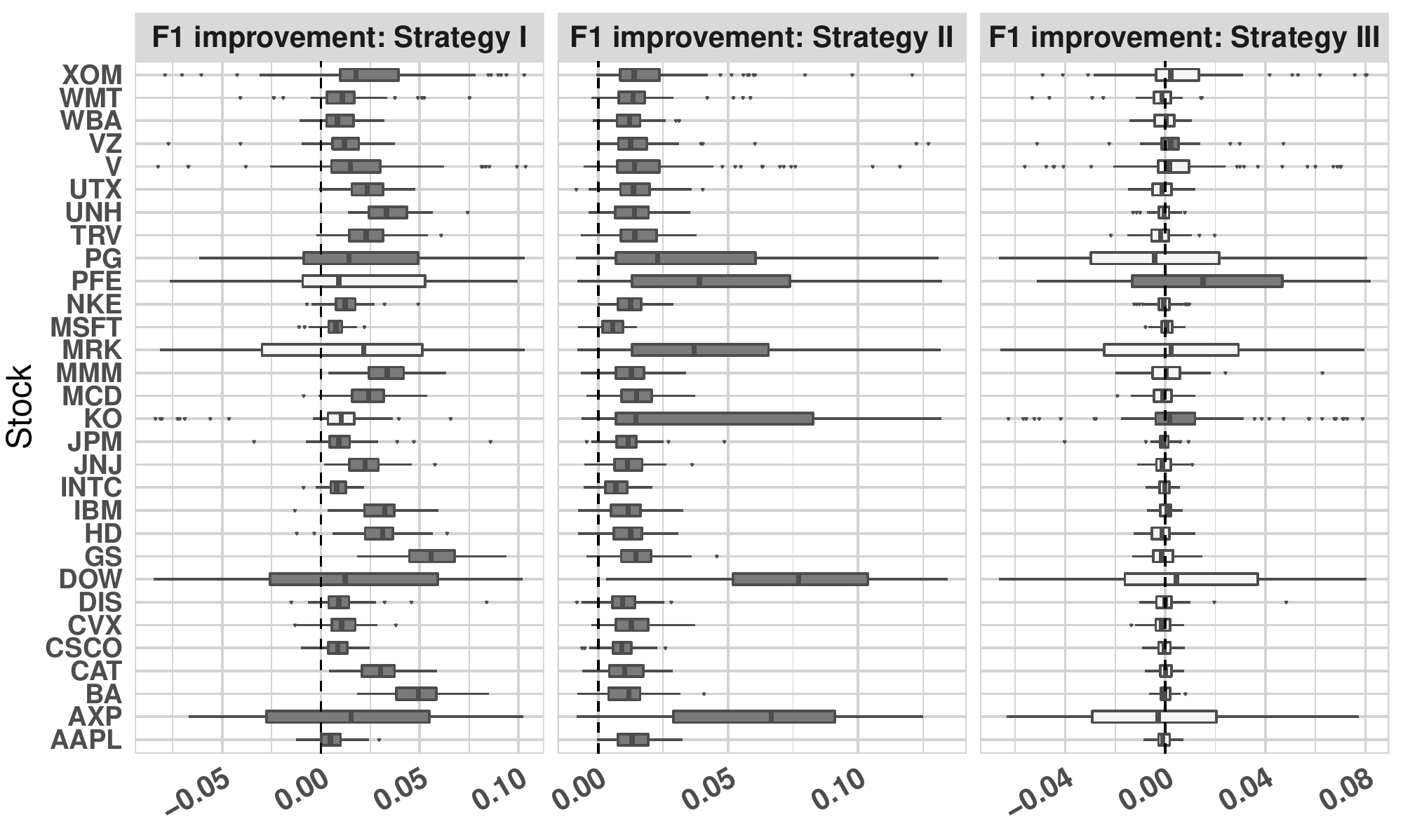}}
\subfigure[Results with utlization of the ENet model\label{Figure7}]{
\includegraphics[width=120mm]{ 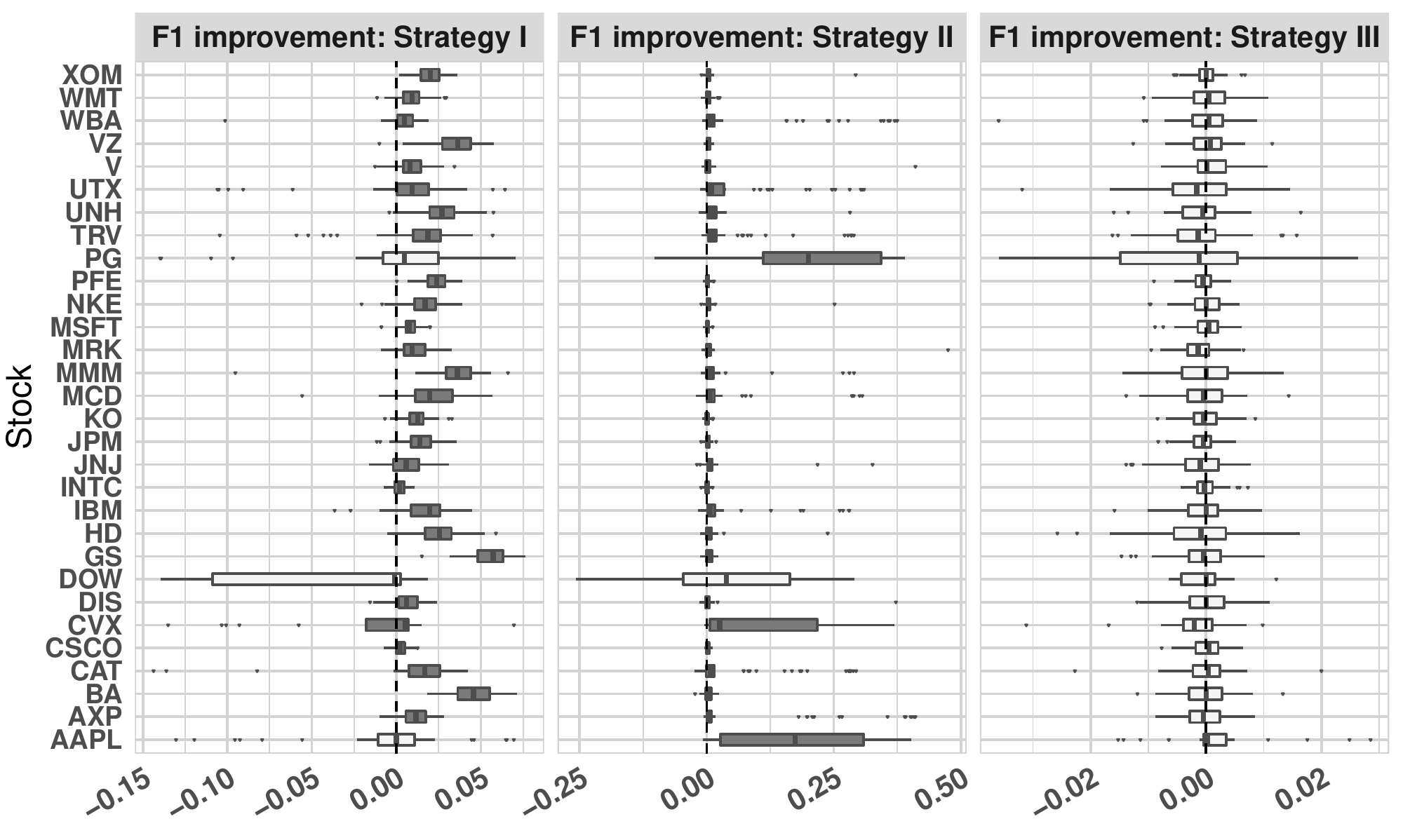}}
\caption{The boxplots show F1 score improvements made by each of our proposed three modelling strategies, when they are used in SVM models (the top panel) and Elastic net models (the bottom model). Each box summarize F1 score improvements in 100 experiments conducted on different random subsets of the full data. A positive value in F1 score change indicates the strategy improve prediction performance. Hence, a box on the right hand side of the vertical dashed line (positioned at zero) indicate proposed strategy is helpful. To inference the significancy of improvement represented by each box, we calculate the raw p-value of Wilcoxon sign rank test, and applied the false discovery rate adjustment \cite{b51} for multiple testing to avoid inflated Type-I error by multiple tests. The boxes corresponding to small adjusted p-values (less than 0.05) are colored in dark gray, which indicate a strategy significantly improve with prediction of that stock, whereas the light gray boxes represent no significant improvements.}
\label{Fig.main}
\end{figure}

We conducted experiments using the component stocks of Dow Jones 30 from the NYSE data. For each prediction performance criteria (precision,  recall, and F1 scores), we obtained 240,000 scores from combinations of the 2 methods (SVM and Enet), 4 models, 30 stocks, and 100 random repeats. For each setting, the median performance scores of 100 random repeats are provided in Appendix Tables ~\ref{d_tab_3} (SVM models) and ~\ref{d_tab_4} (ENet models). In the remainder of the discussion, we focused on the F1 score, as it is the most popular classification performance criteria used in the machine learning community. In each setting, we take the difference in F1 scores between the baseline model and the remaining three models to evaluate performance improvement based on their corresponding strategies. This led to 180,000 F1 score differences. We visualized these F1 score differences in Figure~\ref{Fig.main}, which comprises six panels. The top three panels show the results of the SVM models, and the bottom three panels show the results of the ENet models. From left to right, the panels show the F1 score improvement by each of the three proposed strategies. The results of the 30 stocks are represented by boxes from top to bottom of each panel, and each box represents 100 F1 score differences obtained from repeated experiments. Positive values in the F1 score differences indicate that the corresponding strategy improves the prediction performance; hence, we named it F1 improvement in the panel titles. The dashed vertical line is positioned at zero, which serves as a boundary to distinguish stocks whose mid-price prediction can be improved by the proposed strategy, i.e., boxes on the right-hand side of the boundary. To infer the significance of the improvement represented by each box, we calculated the raw p-value of the Wilcoxon sign rank test and applied the false discovery rate adjustment \cite{b51} for multiple testing to avoid inflated Type-I error by multiple tests. Appendix Table~\ref{d_tab_5} presents the adjusted p-values corresponding to each box in Figure~\ref{Fig.main}. The boxes corresponding to small adjusted p-values (less than 0.05) are colored dark gray, which indicates that a strategy significantly improves the prediction of that stock, whereas the light gray boxes represent no significant improvements.

With SVM, the average improvement in the F1 score over 30 stocks brought by Strategies I, II, and III are 0.02, 0.018, and 0.00036, respectively. The highest improvements of the three strategies were 0.056 (Strategy I on Stock GS), 0.087 (Strategy II on Stock DOW), and 0.016 (Strategy III on Stock PFE). Likewise, regarding ENet model performance, the average improvement in the F1 score through Strategies I, II, and III are 0.016, 0.019, and 0.00046, respectively. The highest improvements of the three strategies are 0.058 (Strategy I on Stock GS), 0.2 (Strategy II on Stock PG), and 0.026 (Strategy III on Stock PG).

We summarize and visualize the performance of our proposed strategies based on the dark grey boxes observed in Figure~\ref{Fig.main}.  Strategy I (variables of `within-window' trends) significantly improved the prediction performance in 27 out of 30 stocks for both the SVM and ENet models. Ensemble learning based on models fitted on many random subsets (Strategy II) significantly improved prediction performance consistently for all stocks, except that the ENet model has one stock showing a positive but non-significant trend. Note that ENet models are not guaranteed to converge. A substantial portion of the ENet model failed to converge in analyzing stocks PG, DOW, and AAPL, which may explain why we have light-gray boxes in Figure~\ref{Fig.main} for Strategies I and II in the ENet model. Therefore, we conclude that the first two strategies are useful for most applications. In contrast, the FPCA of the one-day historical trading record (Strategy III) only helps the SVM models in the three stocks and shows no help in the rest of the predictions. We find that FPCA features are helpful only when daily historical mid-prices are relatively stable. We suggest that users use Strategy III with caution because it only works in specific situations. Users should test Strategy III on their data with various history lengths (e.g., one week, one day, etc.), and use it only if the FPCA of a certain length history seems helpful for prediction of the data.

Appendix Table~\ref{d_tab_2} shows the median computing times of the ENet and SVM models and their corresponding ensemble versions. We found that ENet models were much shorter than SVM models, especially with regard to the ensemble strategy. Thus, we recommend the ENet model, given that it requires fewer computing resources and does not sacrifice much prediction performance. In real-life applications such as HFT, decision time is critical, which makes ENet models more favorable.

\clearpage

\subsection*{Importance of the Predictors \label{importance_of_predictors}}
\begin{figure}[!ht]
    \centering
    \includegraphics[width=120mm]{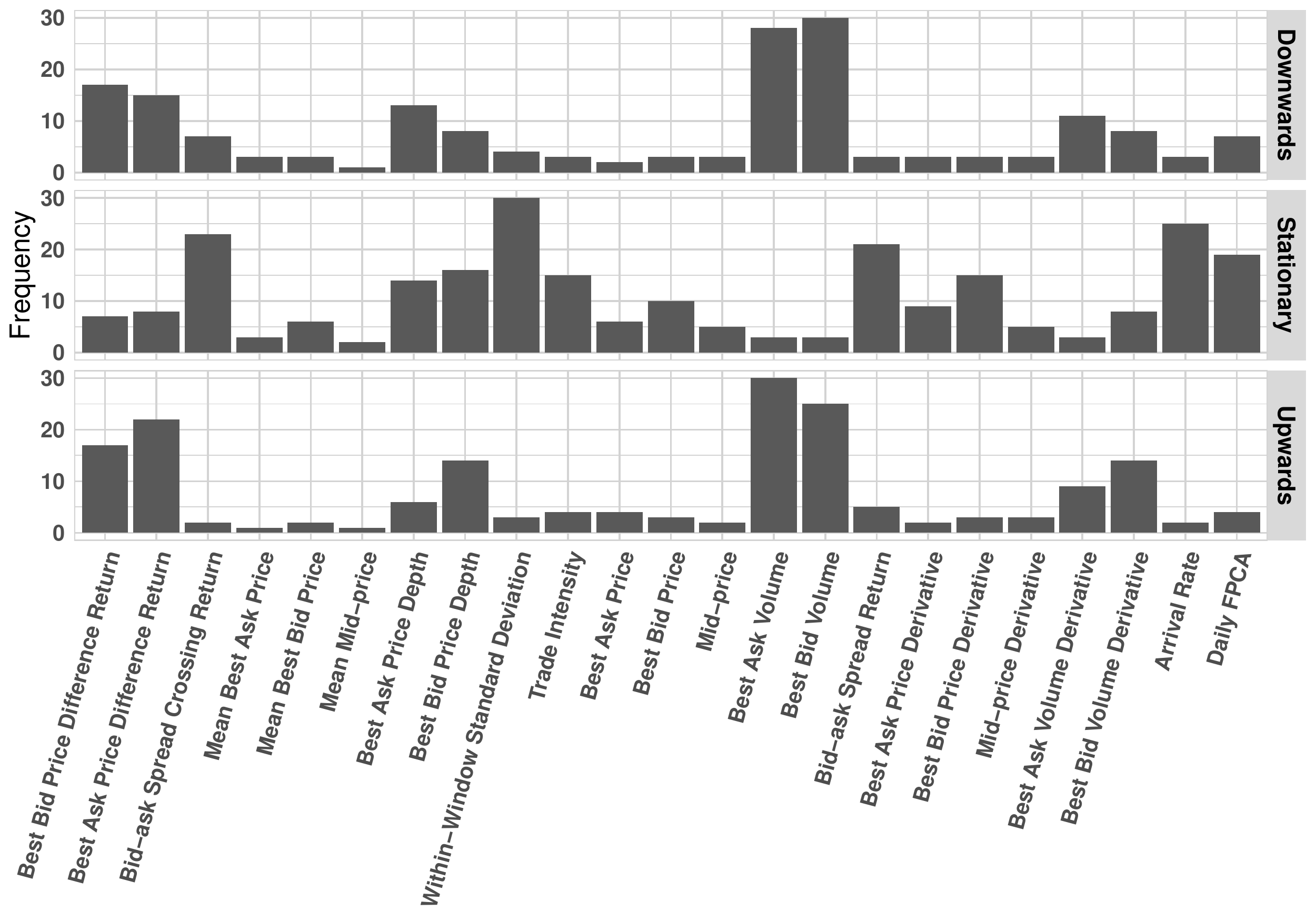}
    \caption{Histogram of total count of high-impact variables of the 30 Dow Jones component stocks in each mid-price movement direction.}
    \label{Figure6}
\end{figure}

The ENet model automatically selects important predictors by assigning zero coefficients to the unimportant predictors. Therefore, we can summarize the importance of the predictors from the above experiment as a by-product. For each stock, we fit numerous ENet models. We consider a predictor to have a high impact if it was selected (i.e., with non-zero coefficients) by $80\%$ of the fitted elastic net models. We believe that the most useful predictors consistently have a high impact on many stocks. Figure~\ref{Figure6} summarizes the frequency of each variable that has a high impact on the 30 component stocks of the Dow Jones 30 index. Because there are 30 stocks, the frequency is in the range $[0,30]$.

From the observed frequencies, we found that for most component stocks, the best bid volume has a high impact in predicting the mid-price movement states of Downwards, the within-window standard deviation has a high impact on predicting the stationary state, and the best asset volume variable has a high impact on predicting the upward state. Many factors, especially from the collection of ``within-window'' high-frequency variables set, are widely chosen to help predict the mid-price movement stationary state, whereas the upwards and downwards states relate more directly to the price differences or the quote volumes from the ask/bid sides. Furthermore, the FPCA scores variables are popular among the prediction of ``stationary'' direction, which confirms that long-term mid-price movement trajectories are useful for predicting stable stock mid-price movement.

\section*{Conclusion \label{conclusion}}

This study proposes three novel strategies to address common issues in predicting high-frequency stock prices using machine learning methods. Our data preprocessing strategies can extract more information from raw data and feed machine learning algorithms with high-quality data input, which is of interest to high-frequency investors. As our first strategy summarizes and introduces the ``within-window'' variables into the model, it recovers the discarded information lost in the event-based inflow protocol during the data thinning process. The second strategy combines a random sampling approach with ensemble machine learning. The sampling method alleviates correlation issues between consecutive observations, while the ensemble method addresses the shortage of potential selection bias caused by random sampling and therefore improves the robustness of the prediction results. Our third strategy sheds light on the effect of long-term trading history on our model. The FPCA reduction of variable dimensionality allows us to model longer-term price curves with few FPCA scores and avoids long vector variables of the sequence data. 

We evaluated the performance of our three proposed strategies using intraday high-frequency trade and quote data from the NYSE and found that Strategies I and II significantly improve prediction performance in most applications. However, Strategy III helps only in certain situations. All three strategies are independent and can be used separately or in combination depending on users' needs. We recommend using Strategies I and II in all applications with high-frequency data that require data-thinning, but only employ Strategy III after testing its performance and carefully exploring the length of history to be utilized in FPCA. Additionally, our strategies are add-ons for use in conjunction with machine-learning models. We illustrate our strategies using SVM and ENet models, and ENet models are preferable because they are computationally faster without sacrificing too much prediction performance.

The proposed method has three limitations. Next, we discuss the study’s limitations and potential solutions. First, Strategy II could be time consuming if excessive ensemble learning is involved, which is problematic in some real-life settings. In cases where the complexity of methods is not linear to the sample size, we may borrow the concept of federated learning \cite{FL, FL2}, in which the model divides data into many smaller samples, learns, and integrates information by updating its parameters. Second, we used FPCA on the hourly resolution to illustrate our strategy, but it might not be the best resolution to reflect the stock's long-term history. We suggest that users explore different resolutions (such as daily or by the minute) and select the best one before applying it to a new stock. The third limitation is that we set model parameters for all stocks using the same rule for illustration purposes, so the performance achieved by an individual stock might not be as ideal as possible. In practice, we recommend that users fine-tune all the relevant model parameters and those in our three strategies for a particular stock. For example, readers can customize any detail in these strategies, which includes the choice of machine learning base learner, the number of trained models to ensemble, the voting scheme in ensemble learning, etc.. Thus, we can obtain the best model performance for each stock.

\section*{Availability of data and materials}
All the processing codes are available upon request.
The data that support the findings of this study are available from the NYSE and TAQ datasets, but restrictions apply to the availability of these data, which were used under license for the current study and so are not publicly available. However, the sample data are available from the authors upon reasonable request.

\clearpage
\section*{Appendix}
\setcounter{table}{0}
\label{app:theorem}

\begin{table}[htbp]
\caption{Components stocks of Dow Jones 30\label{d_tab_1}}
 \centering
\begin{tabular}{lll}
\toprule[1pt]
\textbf{Abbreviation}                & {  \textbf{Company}}        & {  \textbf{Sector}}                \\ \hline
MMM                         & {  3M     }                                     & {  Conglomerate}                   \\
AXP                         & {  American Express}        & {  Financial services}             \\
AAPL                        & {  Apple}                   & {  Information technology}         \\
BA                          & {  Boeing}                  & {  Aerospace and defense}          \\
CAT                         & {  Caterpillar}             & {  Construction and Mining}        \\
CVX                         & {  Chevron}                 & {  Petroleum industry}             \\
CSCO                        & {  Cisco}                   & {  Information technology}         \\
KO                          & {  Coca-Cola}               & {  Food industry}                  \\
DIS                         & {  Disney}                  & {  Broadcasting and entertainment} \\
DOW & {  Dow Chemical}            & {  Chemical industry}              \\
XOM                         & {  Exxon Mobil}             & {  Energy}                         \\
GS                          & {  Goldman Sachs}           & {  Financial services}             \\
HD                          & {  Home Depot}              & {  Retailing}                      \\
IBM                         & {  IBM}                     & {  Information technology}         \\
INTC                        & {  Intel}                   & {  Information technology}         \\
JNJ                         & {  Johnson \& Johnson}      & {  Pharmaceutical industry}        \\
JPM                         & {  JPMorgan Chase}          & {  Financial services}             \\
MCD                         & {  McDonald's}              & {  Food industry}                  \\
MRK                         & {  Merck}                   & {  Pharmaceutical industry}        \\
MSFT                        & {  Microsoft}               & {  Information technology}         \\
NKE                         & {  Nike}                    & {  Apparel}                        \\
PFE                         & {  Pfizer}                  & {  Pharmaceutical industry}        \\
PG                          & {  Procter \& Gamble}       & {  Fast-moving consumer goods}     \\
TRV                         & {  Travelers Companies Inc} & {  Financial services}             \\
UTX                         & {  United Technologies}     & {  Conglomerate}                   \\
UNH                         & {  UnitedHealth}            & {  Managed health care}            \\
VZ                          & {  Verizon}                 & {  Telecommunication}              \\
V                           & {  Visa}                    & {  Financial services}             \\
WMT                         & {  Wal-Mart}                & {  Retailing}                      \\
WBA                         & {  Walgreen}                & {  Retailing}                      \\ 
\toprule[1pt]
\end{tabular}
\end{table}

\begin{table}[ht]
\caption{Median model fitting time (in second) among 100 experiments of pre-defined SVM model and ENet model with 5-folds CV grid search for parameters tuning; median prediction time (in second) among 100 experiments of the ensembled SVM model and ensembled ENet model \label{d_tab_2}}
\centering
\begin{tabular}{rrrrr}
\toprule[1pt]
& SVM & ENet & Ensemble SVM & Ensemble ENet \\
  \hline
AAPL & 15.399 & 64.308 & 9229.379 & 308.853 \\
  MSFT & 409.004 & 62.733 & 6402.511 & 308.500 \\
  MMM & 73.617 & 61.480 & 9448.499 & 321.302 \\
  AXP & 49.939 & 71.618 & 7918.229 & 270.069 \\
  BA & 38.789 & 56.841 & 8553.776 & 304.796 \\
  CAT & 15.442 & 55.063 & 10010.095 & 228.732 \\
  CVX & 14.317 & 69.050 & 10069.264 & 183.757 \\
  CSCO & 487.956 & 65.405 & 7224.831 & 309.895 \\
  KO & 29.856 & 63.751 & 7675.374 & 291.537 \\
  DOW & 163.295 & 74.951 & 8767.993 & 86.906 \\
  XOM & 36.768 & 61.862 & 7506.909 & 198.883 \\
  WBA & 19.297 & 83.645 & 10241.134 & 166.405 \\
  GS & 53.651 & 46.112 & 10038.149 & 315.930 \\
  HD & 18.539 & 54.100 & 10581.459 & 297.982 \\
  INTC & 477.318 & 66.402 & 6714.089 & 321.268 \\
  IBM & 27.033 & 56.349 & 8356.416 & 311.348 \\
  JNJ & 21.766 & 63.325 & 8789.505 & 303.007 \\
  JPM & 22.327 & 56.358 & 7436.894 & 262.190 \\
  MCD & 22.888 & 56.083 & 8704.121 & 183.412 \\
  MRK & 54.528 & 59.175 & 7593.471 & 318.102 \\
  NKE & 23.965 & 64.067 & 7366.485 & 295.981 \\
  PFE & 39.267 & 65.228 & 6652.149 & 323.481 \\
  PG & 94.659 & 86.800 & 7569.769 & 175.051 \\
  TRV & 54.066 & 69.333 & 9042.629 & 296.062 \\
  UNH & 48.844 & 54.428 & 8685.283 & 296.102 \\
  UTX & 35.731 & 65.212 & 8672.759 & 213.724 \\
  VZ & 22.513 & 60.783 & 6900.441 & 295.320 \\
  V & 37.877 & 64.143 & 7624.429 & 275.839 \\
  WMT & 31.197 & 60.831 & 7600.762 & 305.127 \\
  DIS & 14.818 & 62.748 & 9714.119 & 272.908 \\
\toprule[1pt]
\end{tabular}
\end{table}

\begin{table}[ht]
\caption{Median Recall (R), Precision (P) and F1 score (F1) of SVM model for the Dow Jones 30 component stocks among 100 independent experiments under four model setups\label{d_tab_3}}
\resizebox{\textwidth}{70mm}{
\begin{tabular}{ p{1cm}|p{0.9cm}|p{0.9cm}|p{0.9cm}|p{0.9cm}|p{0.9cm}|p{0.9cm}|p{0.9cm}|p{0.9cm}|p{0.9cm}|p{0.9cm}|p{0.9cm}|p{0.9cm}  }
\toprule[1pt]
&\multicolumn{3}{c}{Baseline SVM}
&\multicolumn{3}{c}{Ensemble SVM}&\multicolumn{3}{c}{in-window SVM}&\multicolumn{3}{c}{FPCA SVM}\\
&\multicolumn{3}{c}{(Strategy I, III)}&\multicolumn{3}{c}{(Strategy I, II, III)}&\multicolumn{3}{c}{(Strategy I only)}&\multicolumn{3}{c}{(Strategy III only)}\\
  \hline
& P & R & F1 & P & R & F1 & P & R & F1 & P & R & F1 \\
  \hline
AAPL & 0.572 & 0.563 & 0.567 & 0.586 & 0.577 & 0.581 & 0.572 & 0.563 & 0.567 & 0.573 & 0.553 & 0.562 \\
  MSFT & 0.730 & 0.728 & 0.729 & 0.734 & 0.732 & 0.733 & 0.731 & 0.728 & 0.729 & 0.725 & 0.718 & 0.722 \\
  MMM & 0.457 & 0.448 & 0.451 & 0.467 & 0.462 & 0.465 & 0.457 & 0.448 & 0.452 & 0.422 & 0.414 & 0.419 \\
  AXP & 0.550 & 0.499 & 0.524 & 0.595 & 0.580 & 0.587 & 0.544 & 0.504 & 0.520 & 0.494 & 0.489 & 0.487 \\
  BA & 0.471 & 0.470 & 0.471 & 0.479 & 0.481 & 0.480 & 0.472 & 0.471 & 0.472 & 0.433 & 0.410 & 0.420 \\
  CAT & 0.465 & 0.466 & 0.466 & 0.475 & 0.478 & 0.477 & 0.465 & 0.467 & 0.466 & 0.438 & 0.433 & 0.436 \\
  CVX & 0.522 & 0.518 & 0.520 & 0.533 & 0.531 & 0.532 & 0.524 & 0.518 & 0.521 & 0.518 & 0.501 & 0.509 \\
  CSCO & 0.754 & 0.745 & 0.749 & 0.762 & 0.755 & 0.759 & 0.755 & 0.746 & 0.751 & 0.749 & 0.733 & 0.741 \\
  KO & 0.642 & 0.629 & 0.629 & 0.662 & 0.663 & 0.663 & 0.640 & 0.621 & 0.628 & 0.655 & 0.653 & 0.654 \\
  DOW & 0.504 & 0.486 & 0.494 & 0.584 & 0.589 & 0.586 & 0.494 & 0.484 & 0.491 & 0.488 & 0.478 & 0.480 \\
  XOM & 0.657 & 0.654 & 0.656 & 0.672 & 0.672 & 0.672 & 0.652 & 0.648 & 0.650 & 0.632 & 0.627 & 0.630 \\
  WBA & 0.540 & 0.531 & 0.536 & 0.549 & 0.542 & 0.546 & 0.541 & 0.531 & 0.536 & 0.533 & 0.518 & 0.525 \\
  GS & 0.448 & 0.444 & 0.446 & 0.464 & 0.459 & 0.461 & 0.451 & 0.445 & 0.448 & 0.396 & 0.387 & 0.391 \\
  HD & 0.465 & 0.458 & 0.462 & 0.477 & 0.474 & 0.475 & 0.466 & 0.462 & 0.464 & 0.433 & 0.429 & 0.431 \\
  INTC & 0.738 & 0.727 & 0.732 & 0.744 & 0.733 & 0.739 & 0.738 & 0.727 & 0.732 & 0.732 & 0.716 & 0.724 \\
  IBM & 0.451 & 0.450 & 0.451 & 0.461 & 0.463 & 0.462 & 0.452 & 0.451 & 0.452 & 0.424 & 0.417 & 0.421 \\
  JNJ & 0.511 & 0.512 & 0.512 & 0.524 & 0.526 & 0.525 & 0.513 & 0.513 & 0.513 & 0.492 & 0.488 & 0.489 \\
  JPM & 0.655 & 0.654 & 0.655 & 0.665 & 0.665 & 0.665 & 0.655 & 0.653 & 0.654 & 0.649 & 0.644 & 0.646 \\
  MCD & 0.453 & 0.454 & 0.454 & 0.467 & 0.470 & 0.468 & 0.454 & 0.456 & 0.455 & 0.435 & 0.425 & 0.430 \\
  MRK & 0.439 & 0.381 & 0.406 & 0.484 & 0.385 & 0.429 & 0.425 & 0.365 & 0.397 & 0.436 & 0.363 & 0.398 \\
  NKE & 0.662 & 0.657 & 0.660 & 0.675 & 0.671 & 0.672 & 0.663 & 0.658 & 0.661 & 0.652 & 0.642 & 0.646 \\
  PFE & 0.657 & 0.647 & 0.653 & 0.702 & 0.695 & 0.699 & 0.641 & 0.613 & 0.628 & 0.644 & 0.629 & 0.637 \\
  PG & 0.502 & 0.495 & 0.492 & 0.499 & 0.512 & 0.506 & 0.492 & 0.490 & 0.488 & 0.480 & 0.456 & 0.467 \\
  TRV & 0.437 & 0.434 & 0.435 & 0.450 & 0.451 & 0.451 & 0.439 & 0.435 & 0.438 & 0.417 & 0.409 & 0.412 \\
  UNH & 0.448 & 0.446 & 0.447 & 0.457 & 0.459 & 0.458 & 0.447 & 0.447 & 0.447 & 0.420 & 0.403 & 0.412 \\
  UTX & 0.471 & 0.469 & 0.470 & 0.483 & 0.482 & 0.482 & 0.473 & 0.470 & 0.471 & 0.450 & 0.439 & 0.444 \\
  VZ & 0.703 & 0.701 & 0.702 & 0.716 & 0.717 & 0.716 & 0.703 & 0.700 & 0.701 & 0.692 & 0.688 & 0.690 \\
  V & 0.603 & 0.595 & 0.599 & 0.618 & 0.611 & 0.614 & 0.603 & 0.593 & 0.597 & 0.591 & 0.576 & 0.583 \\
  WMT & 0.632 & 0.625 & 0.629 & 0.643 & 0.637 & 0.640 & 0.635 & 0.626 & 0.631 & 0.623 & 0.613 & 0.618 \\
  DIS & 0.564 & 0.553 & 0.557 & 0.575 & 0.565 & 0.570 & 0.564 & 0.552 & 0.558 & 0.561 & 0.540 & 0.550 \\
\toprule[1pt]
\end{tabular}
}
\end{table}
 
\begin{table}[ht]
\caption{Median Recall (R), Precision (P) and F1 score (F1) of ENet model for the Dow Jones 30 component stocks among 100 independent experiments under four model setups\label{d_tab_4}}
\resizebox{\linewidth}{70mm}{
\begin{tabular}{ p{1cm}|p{0.9cm}|p{0.9cm}|p{0.9cm}|p{0.9cm}|p{0.9cm}|p{0.9cm}|p{0.9cm}|p{0.9cm}|p{0.9cm}|p{0.9cm}|p{0.9cm}|p{0.9cm}  }
\toprule[1pt]
&\multicolumn{3}{c}{Baseline ENet}
&\multicolumn{3}{c}{Ensemble ENet}&\multicolumn{3}{c}{in-window ENet}&\multicolumn{3}{c}{FPCA ENet}\\
&\multicolumn{3}{c}{(Strategy I, III)}&\multicolumn{3}{c}{(Strategy I, II, III)}&\multicolumn{3}{c}{(Strategy I only)}&\multicolumn{3}{c}{(Strategy III only)}\\
  \hline
& P & R & F1 & P & R & F1 & P & R & F1 & P & R & F1 \\
  \hline
AAPL & 0.402 & 0.335 & 0.369 & 0.542 & 0.541 & 0.542 & 0.350 & 0.334 & 0.344 & 0.434 & 0.344 & 0.381 \\
  MSFT & 0.722 & 0.720 & 0.721 & 0.722 & 0.720 & 0.721 & 0.722 & 0.719 & 0.721 & 0.715 & 0.710 & 0.713 \\
  MMM & 0.450 & 0.444 & 0.447 & 0.456 & 0.451 & 0.453 & 0.449 & 0.444 & 0.446 & 0.410 & 0.408 & 0.409 \\
  AXP & 0.561 & 0.559 & 0.560 & 0.569 & 0.568 & 0.569 & 0.562 & 0.558 & 0.559 & 0.555 & 0.553 & 0.554 \\
  BA & 0.464 & 0.459 & 0.461 & 0.468 & 0.462 & 0.466 & 0.464 & 0.459 & 0.461 & 0.418 & 0.414 & 0.416 \\
  CAT & 0.450 & 0.449 & 0.449 & 0.455 & 0.454 & 0.454 & 0.452 & 0.450 & 0.450 & 0.433 & 0.432 & 0.433 \\
  CVX & 0.492 & 0.489 & 0.488 & 0.516 & 0.517 & 0.516 & 0.487 & 0.473 & 0.473 & 0.386 & 0.339 & 0.361 \\
  CSCO & 0.750 & 0.740 & 0.745 & 0.752 & 0.742 & 0.747 & 0.749 & 0.739 & 0.745 & 0.747 & 0.736 & 0.742 \\
  KO & 0.675 & 0.672 & 0.674 & 0.677 & 0.675 & 0.676 & 0.675 & 0.673 & 0.674 & 0.661 & 0.659 & 0.660 \\
  DOW & 0.352 & 0.334 & 0.344 & 0.368 & 0.339 & 0.354 & 0.347 & 0.334 & 0.344 & 0.348 & 0.334 & 0.343 \\
  XOM & 0.644 & 0.644 & 0.644 & 0.649 & 0.647 & 0.648 & 0.645 & 0.645 & 0.645 & 0.623 & 0.626 & 0.625 \\
  WBA & 0.521 & 0.518 & 0.519 & 0.531 & 0.528 & 0.529 & 0.521 & 0.518 & 0.520 & 0.518 & 0.515 & 0.517 \\
  GS & 0.448 & 0.440 & 0.443 & 0.453 & 0.444 & 0.448 & 0.448 & 0.440 & 0.445 & 0.387 & 0.385 & 0.387 \\
  HD & 0.459 & 0.455 & 0.458 & 0.464 & 0.460 & 0.462 & 0.458 & 0.455 & 0.457 & 0.433 & 0.431 & 0.432 \\
  INTC & 0.730 & 0.723 & 0.726 & 0.733 & 0.724 & 0.729 & 0.730 & 0.724 & 0.727 & 0.729 & 0.719 & 0.724 \\
  IBM & 0.439 & 0.436 & 0.437 & 0.446 & 0.444 & 0.445 & 0.439 & 0.436 & 0.437 & 0.422 & 0.418 & 0.420 \\
  JNJ & 0.500 & 0.500 & 0.500 & 0.507 & 0.506 & 0.507 & 0.501 & 0.502 & 0.502 & 0.493 & 0.493 & 0.493 \\
  JPM & 0.631 & 0.628 & 0.630 & 0.634 & 0.633 & 0.633 & 0.630 & 0.629 & 0.629 & 0.614 & 0.614 & 0.614 \\
  MCD & 0.445 & 0.443 & 0.444 & 0.453 & 0.450 & 0.452 & 0.446 & 0.444 & 0.445 & 0.424 & 0.421 & 0.421 \\
  MRK & 0.638 & 0.632 & 0.635 & 0.642 & 0.636 & 0.640 & 0.640 & 0.633 & 0.637 & 0.628 & 0.622 & 0.625 \\
  NKE & 0.640 & 0.636 & 0.638 & 0.644 & 0.641 & 0.642 & 0.639 & 0.636 & 0.638 & 0.623 & 0.619 & 0.621 \\
  PFE & 0.718 & 0.716 & 0.717 & 0.720 & 0.719 & 0.719 & 0.719 & 0.718 & 0.719 & 0.695 & 0.695 & 0.695 \\
  PG & 0.317 & 0.334 & 0.330 & 0.581 & 0.476 & 0.524 & 0.314 & 0.335 & 0.324 & 0.294 & 0.334 & 0.314 \\
  TRV & 0.438 & 0.438 & 0.437 & 0.449 & 0.449 & 0.449 & 0.440 & 0.438 & 0.438 & 0.418 & 0.416 & 0.416 \\
  UNH & 0.439 & 0.437 & 0.439 & 0.448 & 0.448 & 0.448 & 0.440 & 0.440 & 0.440 & 0.411 & 0.409 & 0.410 \\
  UTX & 0.452 & 0.450 & 0.451 & 0.464 & 0.462 & 0.463 & 0.454 & 0.451 & 0.452 & 0.443 & 0.438 & 0.440 \\
  VZ & 0.693 & 0.691 & 0.692 & 0.697 & 0.695 & 0.696 & 0.694 & 0.691 & 0.693 & 0.657 & 0.657 & 0.657 \\
  V & 0.579 & 0.578 & 0.578 & 0.582 & 0.580 & 0.581 & 0.577 & 0.576 & 0.576 & 0.568 & 0.568 & 0.569 \\
  WMT & 0.604 & 0.602 & 0.603 & 0.608 & 0.605 & 0.606 & 0.605 & 0.603 & 0.604 & 0.595 & 0.592 & 0.593 \\
  DIS & 0.540 & 0.540 & 0.539 & 0.543 & 0.541 & 0.542 & 0.541 & 0.540 & 0.541 & 0.533 & 0.533 & 0.533 \\
\toprule[1pt]
\end{tabular}}
\end{table}

\begin{table}[ht]
\caption{Summary of fdr adjusted p-values of three modelling comparison groups with SVM model (left) and ENet model (right) as the base learner respectively \label{d_tab_5}}
\resizebox{\linewidth}{70mm}{
\centering
\begin{tabular}{lrrrrrrr}
\toprule[1pt]
\multicolumn{1}{c}{}&\multicolumn{3}{c}{\textbf{F1 Improvement (SVM)}}
&\multicolumn{3}{c}{\textbf{F1 Improvement (ENet)}}\\
& Strategy I & Strategy II & Strategy III&  Strategy I & Strategy II & Strategy III \\
\hline
AAPL & 1.11e-07 & 1.12e-17 & 1.00e+00 & 7.46e-01 & 2.31e-15 & 1.00e+00 \\
  MSFT & 5.35e-15 & 1.01e-13 & 2.52e-01 & 1.16e-16 & 7.14e-04 & 6.56e-01 \\
  MMM & 7.79e-18 & 1.69e-16 & 1.00e+00 & 2.55e-13 & 2.01e-08 & 1.00e+00 \\
  AXP & 7.51e-06 & 2.90e-17 & 1.00e+00 & 1.03e-03 & 9.29e-10 & 1.00e+00 \\
  BA & 7.79e-18 & 4.01e-16 & 1.00e+00 & 2.89e-17 & 1.44e-05 & 1.00e+00 \\
  CAT & 7.79e-18 & 4.01e-16 & 1.00e+00 & 1.31e-06 & 4.21e-06 & 1.00e+00 \\
  CVX & 2.74e-14 & 1.98e-17 & 1.00e+00 & 4.47e-02 & 1.70e-11 & 1.00e+00 \\
  CSCO & 3.67e-16 & 6.39e-17 & 1.00e+00 & 1.88e-08 & 2.16e-05 & 1.00e+00 \\
  KO & 9.98e-01 & 4.63e-17 & 2.89e-02 & 1.36e-16 & 8.07e-03 & 1.00e+00 \\
  DOW & 3.50e-03 & 1.12e-17 & 1.00e+00 & 9.39e-02 & 1.30e-01 & 1.00e+00 \\
  XOM & 2.14e-12 & 1.12e-17 & 3.16e-01 & 6.79e-13 & 5.73e-07 & 1.00e+00 \\
  WBA & 4.51e-14 & 1.41e-17 & 1.00e+00 & 2.38e-03 & 1.92e-08 & 1.00e+00 \\
  GS & 7.79e-18 & 1.41e-17 & 1.00e+00 & 2.89e-17 & 2.66e-09 & 1.00e+00 \\
  HD & 7.79e-18 & 8.15e-17 & 1.00e+00 & 9.56e-16 & 2.95e-07 & 1.00e+00 \\
  INTC & 6.06e-17 & 3.37e-16 & 1.00e+00 & 2.90e-05 & 3.67e-03 & 1.00e+00 \\
  IBM & 7.79e-18 & 2.27e-16 & 8.96e-01 & 2.35e-11 & 2.01e-11 & 1.00e+00 \\
  JNJ & 7.79e-18 & 2.90e-17 & 1.00e+00 & 3.94e-06 & 4.92e-10 & 1.00e+00 \\
  JPM & 5.70e-16 & 1.41e-17 & 1.00e+00 & 3.56e-15 & 1.21e-05 & 1.00e+00 \\
  MCD & 8.23e-18 & 1.41e-17 & 1.00e+00 & 6.21e-09 & 2.16e-06 & 1.00e+00 \\
  MRK & 5.78e-02 & 3.99e-03 & 4.27e-01 & 3.56e-15 & 4.04e-08 & 1.00e+00 \\
  NKE & 2.72e-17 & 1.12e-17 & 1.00e+00 & 1.23e-14 & 1.92e-08 & 1.00e+00 \\
  PFE & 8.72e-02 & 2.43e-16 & 3.83e-02 & 2.89e-17 & 8.66e-06 & 1.00e+00 \\
  PG & 3.62e-04 & 9.02e-04 & 1.00e+00 & 5.86e-01 & 2.01e-11 & 7.50e-01 \\
  TRV & 7.79e-18 & 2.90e-17 & 1.00e+00 & 2.58e-06 & 1.87e-13 & 1.00e+00 \\
  UNH & 7.79e-18 & 8.15e-17 & 1.00e+00 & 2.63e-15 & 1.70e-10 & 1.00e+00 \\
  UTX & 7.79e-18 & 1.59e-17 & 1.00e+00 & 3.10e-04 & 1.26e-09 & 1.00e+00 \\
  VZ & 2.93e-12 & 1.12e-17 & 2.89e-02 & 5.26e-17 & 1.03e-09 & 7.50e-01 \\
  V & 2.13e-11 & 1.41e-17 & 1.07e-01 & 9.11e-14 & 5.15e-04 & 6.56e-01 \\
  WMT & 3.49e-12 & 1.12e-17 & 1.00e+00 & 2.01e-14 & 8.66e-06 & 6.56e-01 \\
  DIS & 6.80e-15 & 1.32e-15 & 1.00e+00 & 1.20e-08 & 3.11e-03 & 1.00e+00 \\
\toprule[1pt]
\end{tabular}}
\end{table}
\clearpage


\begin{backmatter}

\section*{Abbreviations}
HFT: High-frequency trading; LOB: limit order book; NYSE: New York Stock Exchange; TAQ: Trade and Quote; CV: cross-validation; PCA: Principal component analysis; FPCA: Functional data analysis; FPC: functional principal component; OLS: ordinary least square; IQR: interquartile range; ENet: Elastic net model; SVM: Supporting vector machine; R: Recall; P: Precision; F1: F1 score; MMM:3M; AXP: American Express; AAPL: Apple; BA: Boeing; CAT: Caterpillar; CVX: Chevron; CSCO: Cisco; KO: Coca-Cola; DIS: Disney; DOW: Dow Chemical; XOM: Exxon Mobil; GS: Goldman Sachs; HD: Home Depot; IBM: IBM; INTC: Intel; JNJ: Johnson \& Johnson; JPM: JPMorgan Chase; MCD: McDonald's; MRK: Merck; MSFT: Microsoft; NKE: Nike; PFE: Pfizer; PG: Procter \& Gamble; TRV: Travelers Companies Inc; UTX: United Technologies; UNH: UnitedHealth; VZ: Verizon; V: Visa; WMT: Wal-Mart; WBA: Walgreen.

\section*{Competing interests}
  The authors declare that they have no competing interests.
  
\section*{Funding}
Canada Research Chair (950231363, XZ), Natural Sciences and Engineering Research Council of Canada (NSERC) Discovery Grants (RGPIN-2021–03530, LX), and the Social Sciences and Humanities Research Council of Canada (SSHRC) Insight Development Grants (430-2018-00557, KX).

\section*{Author's contributions}
XZ supervised this project. XZ and LX contributed to the conceptualization and design of the study. YH developed computer programs and conducted the experiments. KX provided data and supported modelling and the interpretation of the results' financial meaning. XZ and YH contributed to the manuscript preparation, and all authors contributed to the revision and approved the final draft.

\section*{Acknowledgements}
  The authors acknowledge that this research was enabled in part by support provided by WestGrid (www.westgrid.ca) and Compute Canada (www.computecanada.ca).


\bibliographystyle{bmc-mathphys} 
\bibliography{bmc_article}      


\begin{thebibliography}{32}
\ifx \bisbn   \undefined \def \bisbn  #1{ISBN #1}\fi
\ifx \binits  \undefined \def \binits#1{#1}\fi
\ifx \bauthor  \undefined \def \bauthor#1{#1}\fi
\ifx \batitle  \undefined \def \batitle#1{#1}\fi
\ifx \bjtitle  \undefined \def \bjtitle#1{#1}\fi
\ifx \bvolume  \undefined \def \bvolume#1{\textbf{#1}}\fi
\ifx \byear  \undefined \def \byear#1{#1}\fi
\ifx \bissue  \undefined \def \bissue#1{#1}\fi
\ifx \bfpage  \undefined \def \bfpage#1{#1}\fi
\ifx \blpage  \undefined \def \blpage #1{#1}\fi
\ifx \burl  \undefined \def \burl#1{\textsf{#1}}\fi
\ifx \doiurl  \undefined \def \doiurl#1{\textsf{#1}}\fi
\ifx \betal  \undefined \def \betal{\textit{et al.}}\fi
\ifx \binstitute  \undefined \def \binstitute#1{#1}\fi
\ifx \binstitutionaled  \undefined \def \binstitutionaled#1{#1}\fi
\ifx \bctitle  \undefined \def \bctitle#1{#1}\fi
\ifx \beditor  \undefined \def \beditor#1{#1}\fi
\ifx \bpublisher  \undefined \def \bpublisher#1{#1}\fi
\ifx \bbtitle  \undefined \def \bbtitle#1{#1}\fi
\ifx \bedition  \undefined \def \bedition#1{#1}\fi
\ifx \bseriesno  \undefined \def \bseriesno#1{#1}\fi
\ifx \blocation  \undefined \def \blocation#1{#1}\fi
\ifx \bsertitle  \undefined \def \bsertitle#1{#1}\fi
\ifx \bsnm \undefined \def \bsnm#1{#1}\fi
\ifx \bsuffix \undefined \def \bsuffix#1{#1}\fi
\ifx \bparticle \undefined \def \bparticle#1{#1}\fi
\ifx \barticle \undefined \def \barticle#1{#1}\fi
\ifx \bconfdate \undefined \def \bconfdate #1{#1}\fi
\ifx \botherref \undefined \def \botherref #1{#1}\fi
\ifx \url \undefined \def \url#1{\textsf{#1}}\fi
\ifx \bchapter \undefined \def \bchapter#1{#1}\fi
\ifx \bbook \undefined \def \bbook#1{#1}\fi
\ifx \bcomment \undefined \def \bcomment#1{#1}\fi
\ifx \oauthor \undefined \def \oauthor#1{#1}\fi
\ifx \citeauthoryear \undefined \def \citeauthoryear#1{#1}\fi
\ifx \endbibitem  \undefined \def \endbibitem {}\fi
\ifx \bconflocation  \undefined \def \bconflocation#1{#1}\fi
\ifx \arxivurl  \undefined \def \arxivurl#1{\textsf{#1}}\fi
\csname PreBibitemsHook\endcsname

\bibitem{b1}
\begin{barticle}
\bauthor{\bsnm{Securities}},
\bauthor{\bsnm{Commission}, \binits{E.}}:
\batitle{Concept release on equity market structure}.
\bjtitle{IEEE Transactions on Information Theory}
\bvolume{34}(\bissue{61358}),
\bfpage{7}--\blpage{0210}
(\byear{2010})
\end{barticle}
\endbibitem

\bibitem{b2}
\begin{barticle}
\bauthor{\bsnm{Menkveld}, \binits{A.J.}}:
\batitle{High frequency trading and the new market makers}.
\bjtitle{Journal of financial Markets}
\bvolume{16}(\bissue{4}),
\bfpage{712}--\blpage{740}
(\byear{2013})
\end{barticle}
\endbibitem

\bibitem{b3}
\begin{barticle}
\bauthor{\bsnm{Parlour}, \binits{C.A.}},
\bauthor{\bsnm{Seppi}, \binits{D.J.}}:
\batitle{Limit order markets: A survey}.
\bjtitle{Handbook of financial intermediation and banking}
\bvolume{5},
\bfpage{63}--\blpage{95}
(\byear{2008})
\end{barticle}
\endbibitem

\bibitem{b50}
\begin{botherref}
\oauthor{\bsnm{Qian}, \binits{X.-Y.}},
\oauthor{\bsnm{Gao}, \binits{S.}}:
Financial series prediction: Comparison between precision of time series models
  and machine learning methods.
arXiv preprint arXiv:1706.00948,
1--9
(2017)
\end{botherref}
\endbibitem

\bibitem{li2021integrated}
\begin{botherref}
\oauthor{\bsnm{Li}, \binits{T.}},
\oauthor{\bsnm{Kou}, \binits{G.}},
\oauthor{\bsnm{Peng}, \binits{Y.}},
\oauthor{\bsnm{Philip}, \binits{S.Y.}}:
An integrated cluster detection, optimization, and interpretation approach for
  financial data.
IEEE Transactions on Cybernetics
(2021)
\end{botherref}
\endbibitem

\bibitem{b7}
\begin{botherref}
\oauthor{\bsnm{Qiao}, \binits{Q.}},
\oauthor{\bsnm{Beling}, \binits{P.A.}}:
Decision analytics and machine learning in economic and financial systems.
Springer
(2016)
\end{botherref}
\endbibitem

\bibitem{huang2017nonlinear}
\begin{barticle}
\bauthor{\bsnm{Huang}, \binits{Y.}},
\bauthor{\bsnm{Kou}, \binits{G.}},
\bauthor{\bsnm{Peng}, \binits{Y.}}:
\batitle{Nonlinear manifold learning for early warnings in financial markets}.
\bjtitle{European Journal of Operational Research}
\bvolume{258}(\bissue{2}),
\bfpage{692}--\blpage{702}
(\byear{2017})
\end{barticle}
\endbibitem

\bibitem{b10}
\begin{barticle}
\bauthor{\bsnm{Chen}, \binits{A.-S.}},
\bauthor{\bsnm{Leung}, \binits{M.T.}},
\bauthor{\bsnm{Daouk}, \binits{H.}}:
\batitle{Application of neural networks to an emerging financial market:
  forecasting and trading the taiwan stock index}.
\bjtitle{Computers \& Operations Research}
\bvolume{30}(\bissue{6}),
\bfpage{901}--\blpage{923}
(\byear{2003})
\end{barticle}
\endbibitem

\bibitem{wen2019retail}
\begin{barticle}
\bauthor{\bsnm{Wen}, \binits{F.}},
\bauthor{\bsnm{Xu}, \binits{L.}},
\bauthor{\bsnm{Ouyang}, \binits{G.}},
\bauthor{\bsnm{Kou}, \binits{G.}}:
\batitle{Retail investor attention and stock price crash risk: evidence from
  china}.
\bjtitle{International Review of Financial Analysis}
\bvolume{65},
\bfpage{101376}
(\byear{2019})
\end{barticle}
\endbibitem

\bibitem{b11}
\begin{barticle}
\bauthor{\bsnm{Fletcher}, \binits{T.}},
\bauthor{\bsnm{Shawe-Taylor}, \binits{J.}}:
\batitle{Multiple kernel learning with fisher kernels for high frequency
  currency prediction}.
\bjtitle{Computational Economics}
\bvolume{42}(\bissue{2}),
\bfpage{217}--\blpage{240}
(\byear{2013})
\end{barticle}
\endbibitem

\bibitem{b12}
\begin{barticle}
\bauthor{\bsnm{Kercheval}, \binits{A.N.}},
\bauthor{\bsnm{Zhang}, \binits{Y.}}:
\batitle{Modelling high-frequency limit order book dynamics with support vector
  machines}.
\bjtitle{Quantitative Finance}
\bvolume{15}(\bissue{8}),
\bfpage{1315}--\blpage{1329}
(\byear{2015})
\end{barticle}
\endbibitem

\bibitem{b44}
\begin{bchapter}
\bauthor{\bsnm{Ar{\'e}valo}, \binits{A.}},
\bauthor{\bsnm{Ni{\~n}o}, \binits{J.}},
\bauthor{\bsnm{Hern{\'a}ndez}, \binits{G.}},
\bauthor{\bsnm{Sandoval}, \binits{J.}}:
\bctitle{High-frequency trading strategy based on deep neural networks}.
In: \bbtitle{International Conference on Intelligent Computing},
pp. \bfpage{424}--\blpage{436}
(\byear{2016}).
\bcomment{Springer}
\end{bchapter}
\endbibitem

\bibitem{b45}
\begin{botherref}
\oauthor{\bsnm{Dixon}, \binits{M.}}:
High frequency market making with machine learning.
November
(2016)
\end{botherref}
\endbibitem

\bibitem{b46}
\begin{barticle}
\bauthor{\bsnm{Kong}, \binits{A.}},
\bauthor{\bsnm{Zhu}, \binits{H.}}:
\batitle{Predicting trend of high frequency csi 300 index using adaptive input
  selection and machine learning techniques}.
\bjtitle{Journal of Systems Science and Information}
\bvolume{6}(\bissue{2}),
\bfpage{120}--\blpage{133}
(\byear{2018})
\end{barticle}
\endbibitem

\bibitem{b47}
\begin{botherref}
\oauthor{\bsnm{Campbell}, \binits{J.Y.}},
\oauthor{\bsnm{Grossman}, \binits{S.J.}},
\oauthor{\bsnm{Wang}, \binits{J.}}:
{Trading Volume and Serial Correlation in Stock Returns}.
NBER Working Papers 4193,
National Bureau of Economic Research, Inc
(October 1992).
\url{https://ideas.repec.org/p/nbr/nberwo/4193.html}
\end{botherref}
\endbibitem

\bibitem{b48}
\begin{bbook}
\bauthor{\bsnm{Campbell}, \binits{J.Y.}},
\bauthor{\bsnm{Lo}, \binits{A.W.}},
\bauthor{\bsnm{MacKinlay}, \binits{A.C.}}:
\bbtitle{The Econometrics of Financial Markets}.
\bpublisher{Princeton University Press},
\blocation{Princeton, New Jersey}
(\byear{2012})
\end{bbook}
\endbibitem

\bibitem{b38}
\begin{barticle}
\bauthor{\bsnm{Ntakaris}, \binits{A.}},
\bauthor{\bsnm{Magris}, \binits{M.}},
\bauthor{\bsnm{Kanniainen}, \binits{J.}},
\bauthor{\bsnm{Gabbouj}, \binits{M.}},
\bauthor{\bsnm{Iosifidis}, \binits{A.}}:
\batitle{Benchmark dataset for mid-price forecasting of limit order book data
  with machine learning methods}.
\bjtitle{Journal of Forecasting}
\bvolume{37}(\bissue{8}),
\bfpage{852}--\blpage{866}
(\byear{2018})
\end{barticle}
\endbibitem

\bibitem{b39}
\begin{barticle}
\bauthor{\bsnm{Nousi}, \binits{P.}},
\bauthor{\bsnm{Tsantekidis}, \binits{A.}},
\bauthor{\bsnm{Passalis}, \binits{N.}},
\bauthor{\bsnm{Ntakaris}, \binits{A.}},
\bauthor{\bsnm{Kanniainen}, \binits{J.}},
\bauthor{\bsnm{Tefas}, \binits{A.}},
\bauthor{\bsnm{Gabbouj}, \binits{M.}},
\bauthor{\bsnm{Iosifidis}, \binits{A.}}:
\batitle{Machine learning for forecasting mid-price movements using limit order
  book data}.
\bjtitle{Ieee Access}
\bvolume{7},
\bfpage{64722}--\blpage{64736}
(\byear{2019})
\end{barticle}
\endbibitem

\bibitem{b26}
\begin{barticle}
\bauthor{\bsnm{Ntakaris}, \binits{A.}},
\bauthor{\bsnm{Mirone}, \binits{G.}},
\bauthor{\bsnm{Kanniainen}, \binits{J.}},
\bauthor{\bsnm{Gabbouj}, \binits{M.}},
\bauthor{\bsnm{Iosifidis}, \binits{A.}}:
\batitle{Feature engineering for mid-price prediction with deep learning}.
\bjtitle{Ieee Access}
\bvolume{7},
\bfpage{82390}--\blpage{82412}
(\byear{2019})
\end{barticle}
\endbibitem

\bibitem{b49}
\begin{barticle}
\bauthor{\bsnm{Ntakaris}, \binits{A.}},
\bauthor{\bsnm{Kanniainen}, \binits{J.}},
\bauthor{\bsnm{Gabbouj}, \binits{M.}},
\bauthor{\bsnm{Iosifidis}, \binits{A.}}:
\batitle{Mid-price prediction based on machine learning methods with technical
  and quantitative indicators}.
\bjtitle{Plos one}
\bvolume{15}(\bissue{6}),
\bfpage{0234107}
(\byear{2020})
\end{barticle}
\endbibitem

\bibitem{b40}
\begin{botherref}
\oauthor{\bsnm{Ramsay}, \binits{J.O.}}:
Functional data analysis.
Encyclopedia of Statistical Sciences
\textbf{4}
(2004)
\end{botherref}
\endbibitem

\bibitem{b41}
\begin{bbook}
\bauthor{\bsnm{Ramsay}, \binits{J.O.}},
\bauthor{\bsnm{Silverman}, \binits{B.W.}}:
\bbtitle{Applied Functional Data Analysis: Methods and Case Studies}.
\bpublisher{Springer},
\blocation{Germany}
(\byear{2007})
\end{bbook}
\endbibitem

\bibitem{b42}
\begin{bbook}
\bauthor{\bsnm{Kokoszka}, \binits{P.}},
\bauthor{\bsnm{Reimherr}, \binits{M.}}:
\bbtitle{Introduction to Functional Data Analysis}.
\bpublisher{CRC press},
\blocation{Boca Raton}
(\byear{2017})
\end{bbook}
\endbibitem

\bibitem{b17}
\begin{barticle}
\bauthor{\bsnm{Tay}, \binits{F.E.}},
\bauthor{\bsnm{Cao}, \binits{L.}}:
\batitle{Application of support vector machines in financial time series
  forecasting}.
\bjtitle{omega}
\bvolume{29}(\bissue{4}),
\bfpage{309}--\blpage{317}
(\byear{2001})
\end{barticle}
\endbibitem

\bibitem{b18}
\begin{barticle}
\bauthor{\bsnm{Huang}, \binits{W.}},
\bauthor{\bsnm{Nakamori}, \binits{Y.}},
\bauthor{\bsnm{Wang}, \binits{S.-Y.}}:
\batitle{Forecasting stock market movement direction with support vector
  machine}.
\bjtitle{Computers \& operations research}
\bvolume{32}(\bissue{10}),
\bfpage{2513}--\blpage{2522}
(\byear{2005})
\end{barticle}
\endbibitem

\bibitem{b19}
\begin{bchapter}
\bauthor{\bsnm{Chalup}, \binits{S.K.}},
\bauthor{\bsnm{Mitschele}, \binits{A.}}:
\bctitle{Kernel methods in finance}.
In: \bbtitle{Handbook on Information Technology in Finance},
pp. \bfpage{655}--\blpage{687}.
\bpublisher{Springer},
\blocation{Germany}
(\byear{2008})
\end{bchapter}
\endbibitem

\bibitem{b24}
\begin{barticle}
\bauthor{\bsnm{Zou}, \binits{H.}},
\bauthor{\bsnm{Hastie}, \binits{T.}}:
\batitle{Regularization and variable selection via the elastic net}.
\bjtitle{Journal of the royal statistical society: series B (statistical
  methodology)}
\bvolume{67}(\bissue{2}),
\bfpage{301}--\blpage{320}
(\byear{2005})
\end{barticle}
\endbibitem

\bibitem{b27}
\begin{barticle}
\bauthor{\bsnm{Hendershott}, \binits{T.}},
\bauthor{\bsnm{Moulton}, \binits{P.C.}}:
\batitle{Automation, speed, and stock market quality: The nyse's hybrid}.
\bjtitle{Journal of Financial Markets}
\bvolume{14}(\bissue{4}),
\bfpage{568}--\blpage{604}
(\byear{2011})
\end{barticle}
\endbibitem

\bibitem{b43}
\begin{barticle}
\bauthor{\bsnm{Friedman}, \binits{J.}},
\bauthor{\bsnm{Hastie}, \binits{T.}},
\bauthor{\bsnm{Tibshirani}, \binits{R.}}:
\batitle{Regularization paths for generalized linear models via coordinate
  descent}.
\bjtitle{Journal of Statistical Software}
\bvolume{33}(\bissue{1}),
\bfpage{1}--\blpage{22}
(\byear{2010})
\end{barticle}
\endbibitem

\bibitem{b51}
\begin{barticle}
\bauthor{\bsnm{Benjamini}, \binits{Y.}},
\bauthor{\bsnm{Hochberg}, \binits{Y.}}:
\batitle{Controlling the false discovery rate: a practical and powerful
  approach to multiple testing}.
\bjtitle{Journal of the Royal statistical society: series B (Methodological)}
\bvolume{57}(\bissue{1}),
\bfpage{289}--\blpage{300}
(\byear{1995})
\end{barticle}
\endbibitem

\bibitem{FL}
\begin{barticle}
\bauthor{\bsnm{Li}, \binits{T.}},
\bauthor{\bsnm{Sahu}, \binits{A.K.}},
\bauthor{\bsnm{Talwalkar}, \binits{A.}},
\bauthor{\bsnm{Smith}, \binits{V.}}:
\batitle{Federated learning: Challenges, methods, and future directions}.
\bjtitle{IEEE Signal Processing Magazine}
\bvolume{37}(\bissue{3}),
\bfpage{50}--\blpage{60}
(\byear{2020}).
doi:\doiurl{10.1109/MSP.2020.2975749}
\end{barticle}
\endbibitem

\bibitem{FL2}
\begin{barticle}
\bauthor{\bsnm{Kairouz}, \binits{P.}},
\bauthor{\bsnm{McMahan}, \binits{H.B.}},
\bauthor{\bsnm{Avent}, \binits{B.}},
\bauthor{\bsnm{Bellet}, \binits{A.}},
\bauthor{\bsnm{Bennis}, \binits{M.}},
\bauthor{\bsnm{Bhagoji}, \binits{A.N.}},
\bauthor{\bsnm{Bonawitz}, \binits{K.}},
\bauthor{\bsnm{Charles}, \binits{Z.}},
\bauthor{\bsnm{Cormode}, \binits{G.}},
\bauthor{\bsnm{Cummings}, \binits{R.}}, \betal:
\batitle{Advances and open problems in federated learning}.
\bjtitle{Foundations and Trends{\textregistered} in Machine Learning}
\bvolume{14}(\bissue{1--2}),
\bfpage{1}--\blpage{210}
(\byear{2021})
\end{barticle}
\endbibitem

\end{thebibliography}

\newcommand{\BMCxmlcomment}[1]{}

\BMCxmlcomment{

<refgrp>

<bibl id="B1">
  <title><p>Concept Release on Equity Market Structure</p></title>
  <aug>
    <au><cnm>Securities</cnm></au>
    <au><snm>Commission</snm><fnm>E</fnm></au>
  </aug>
  <source>IEEE Transactions on Information Theory</source>
  <pubdate>2010</pubdate>
  <volume>34</volume>
  <issue>61358</issue>
  <fpage>S7</fpage>
  <lpage>02-10</lpage>
</bibl>

<bibl id="B2">
  <title><p>High frequency trading and the new market makers</p></title>
  <aug>
    <au><snm>Menkveld</snm><fnm>AJ</fnm></au>
  </aug>
  <source>Journal of financial Markets</source>
  <publisher>Elsevier</publisher>
  <pubdate>2013</pubdate>
  <volume>16</volume>
  <issue>4</issue>
  <fpage>712</fpage>
  <lpage>-740</lpage>
</bibl>

<bibl id="B3">
  <title><p>Limit order markets: A survey</p></title>
  <aug>
    <au><snm>Parlour</snm><fnm>CA</fnm></au>
    <au><snm>Seppi</snm><fnm>DJ</fnm></au>
  </aug>
  <source>Handbook of financial intermediation and banking</source>
  <publisher>Citeseer</publisher>
  <pubdate>2008</pubdate>
  <volume>5</volume>
  <fpage>63</fpage>
  <lpage>-95</lpage>
</bibl>

<bibl id="B4">
  <title><p>Financial series prediction: Comparison between precision of time
  series models and machine learning methods</p></title>
  <aug>
    <au><snm>Qian</snm><fnm>XY</fnm></au>
    <au><snm>Gao</snm><fnm>S</fnm></au>
  </aug>
  <source>arXiv preprint arXiv:1706.00948</source>
  <pubdate>2017</pubdate>
  <fpage>1</fpage>
  <lpage>-9</lpage>
</bibl>

<bibl id="B5">
  <title><p>An integrated cluster detection, optimization, and interpretation
  approach for financial data</p></title>
  <aug>
    <au><snm>Li</snm><fnm>T</fnm></au>
    <au><snm>Kou</snm><fnm>G</fnm></au>
    <au><snm>Peng</snm><fnm>Y</fnm></au>
    <au><snm>Philip</snm><fnm>SY</fnm></au>
  </aug>
  <source>IEEE Transactions on Cybernetics</source>
  <publisher>IEEE</publisher>
  <pubdate>2021</pubdate>
</bibl>

<bibl id="B6">
  <title><p>Decision analytics and machine learning in economic and financial
  systems</p></title>
  <aug>
    <au><snm>Qiao</snm><fnm>Q</fnm></au>
    <au><snm>Beling</snm><fnm>PA</fnm></au>
  </aug>
  <publisher>Springer</publisher>
  <pubdate>2016</pubdate>
</bibl>

<bibl id="B7">
  <title><p>Nonlinear manifold learning for early warnings in financial
  markets</p></title>
  <aug>
    <au><snm>Huang</snm><fnm>Y</fnm></au>
    <au><snm>Kou</snm><fnm>G</fnm></au>
    <au><snm>Peng</snm><fnm>Y</fnm></au>
  </aug>
  <source>European Journal of Operational Research</source>
  <publisher>Elsevier</publisher>
  <pubdate>2017</pubdate>
  <volume>258</volume>
  <issue>2</issue>
  <fpage>692</fpage>
  <lpage>-702</lpage>
</bibl>

<bibl id="B8">
  <title><p>Application of neural networks to an emerging financial market:
  forecasting and trading the Taiwan Stock Index</p></title>
  <aug>
    <au><snm>Chen</snm><fnm>AS</fnm></au>
    <au><snm>Leung</snm><fnm>MT</fnm></au>
    <au><snm>Daouk</snm><fnm>H</fnm></au>
  </aug>
  <source>Computers \& Operations Research</source>
  <publisher>Elsevier</publisher>
  <pubdate>2003</pubdate>
  <volume>30</volume>
  <issue>6</issue>
  <fpage>901</fpage>
  <lpage>-923</lpage>
</bibl>

<bibl id="B9">
  <title><p>Retail investor attention and stock price crash risk: evidence from
  China</p></title>
  <aug>
    <au><snm>Wen</snm><fnm>F</fnm></au>
    <au><snm>Xu</snm><fnm>L</fnm></au>
    <au><snm>Ouyang</snm><fnm>G</fnm></au>
    <au><snm>Kou</snm><fnm>G</fnm></au>
  </aug>
  <source>International Review of Financial Analysis</source>
  <publisher>Elsevier</publisher>
  <pubdate>2019</pubdate>
  <volume>65</volume>
  <fpage>101376</fpage>
</bibl>

<bibl id="B10">
  <title><p>Multiple kernel learning with fisher kernels for high frequency
  currency prediction</p></title>
  <aug>
    <au><snm>Fletcher</snm><fnm>T</fnm></au>
    <au><snm>Shawe Taylor</snm><fnm>J</fnm></au>
  </aug>
  <source>Computational Economics</source>
  <publisher>Springer</publisher>
  <pubdate>2013</pubdate>
  <volume>42</volume>
  <issue>2</issue>
  <fpage>217</fpage>
  <lpage>-240</lpage>
</bibl>

<bibl id="B11">
  <title><p>Modelling high-frequency limit order book dynamics with support
  vector machines</p></title>
  <aug>
    <au><snm>Kercheval</snm><fnm>AN</fnm></au>
    <au><snm>Zhang</snm><fnm>Y</fnm></au>
  </aug>
  <source>Quantitative Finance</source>
  <publisher>Taylor \& Francis</publisher>
  <pubdate>2015</pubdate>
  <volume>15</volume>
  <issue>8</issue>
  <fpage>1315</fpage>
  <lpage>-1329</lpage>
</bibl>

<bibl id="B12">
  <title><p>High-frequency trading strategy based on deep neural
  networks</p></title>
  <aug>
    <au><snm>Ar{\'e}valo</snm><fnm>A</fnm></au>
    <au><snm>Ni{\~n}o</snm><fnm>J</fnm></au>
    <au><snm>Hern{\'a}ndez</snm><fnm>G</fnm></au>
    <au><snm>Sandoval</snm><fnm>J</fnm></au>
  </aug>
  <source>International conference on intelligent computing</source>
  <pubdate>2016</pubdate>
  <fpage>424</fpage>
  <lpage>-436</lpage>
</bibl>

<bibl id="B13">
  <title><p>High frequency market making with machine learning</p></title>
  <aug>
    <au><snm>Dixon</snm><fnm>M</fnm></au>
  </aug>
  <publisher>November</publisher>
  <pubdate>2016</pubdate>
</bibl>

<bibl id="B14">
  <title><p>Predicting trend of high frequency CSI 300 index using adaptive
  input selection and machine learning techniques</p></title>
  <aug>
    <au><snm>Kong</snm><fnm>A</fnm></au>
    <au><snm>Zhu</snm><fnm>H</fnm></au>
  </aug>
  <source>Journal of Systems Science and Information</source>
  <publisher>De Gruyter</publisher>
  <pubdate>2018</pubdate>
  <volume>6</volume>
  <issue>2</issue>
  <fpage>120</fpage>
  <lpage>-133</lpage>
</bibl>

<bibl id="B15">
  <title><p>{Trading Volume and Serial Correlation in Stock
  Returns}</p></title>
  <aug>
    <au><snm>Campbell</snm><fnm>JY</fnm></au>
    <au><snm>Grossman</snm><fnm>SJ</fnm></au>
    <au><snm>Wang</snm><fnm>J</fnm></au>
  </aug>
  <source>NBER Working Papers</source>
  <pubdate>1992</pubdate>
  <issue>4193</issue>
  <url>https://ideas.repec.org/p/nbr/nberwo/4193.html</url>
</bibl>

<bibl id="B16">
  <title><p>The Econometrics of Financial Markets</p></title>
  <aug>
    <au><snm>Campbell</snm><fnm>J.Y.</fnm></au>
    <au><snm>Lo</snm><fnm>A.W.</fnm></au>
    <au><snm>MacKinlay</snm><fnm>A.C.</fnm></au>
  </aug>
  <publisher>Princeton, New Jersey: Princeton University Press</publisher>
  <pubdate>2012</pubdate>
</bibl>

<bibl id="B17">
  <title><p>Benchmark dataset for mid-price forecasting of limit order book
  data with machine learning methods</p></title>
  <aug>
    <au><snm>Ntakaris</snm><fnm>A</fnm></au>
    <au><snm>Magris</snm><fnm>M</fnm></au>
    <au><snm>Kanniainen</snm><fnm>J</fnm></au>
    <au><snm>Gabbouj</snm><fnm>M</fnm></au>
    <au><snm>Iosifidis</snm><fnm>A</fnm></au>
  </aug>
  <source>Journal of Forecasting</source>
  <publisher>Wiley Online Library</publisher>
  <pubdate>2018</pubdate>
  <volume>37</volume>
  <issue>8</issue>
  <fpage>852</fpage>
  <lpage>-866</lpage>
</bibl>

<bibl id="B18">
  <title><p>Machine learning for forecasting mid-price movements using limit
  order book data</p></title>
  <aug>
    <au><snm>Nousi</snm><fnm>P</fnm></au>
    <au><snm>Tsantekidis</snm><fnm>A</fnm></au>
    <au><snm>Passalis</snm><fnm>N</fnm></au>
    <au><snm>Ntakaris</snm><fnm>A</fnm></au>
    <au><snm>Kanniainen</snm><fnm>J</fnm></au>
    <au><snm>Tefas</snm><fnm>A</fnm></au>
    <au><snm>Gabbouj</snm><fnm>M</fnm></au>
    <au><snm>Iosifidis</snm><fnm>A</fnm></au>
  </aug>
  <source>Ieee Access</source>
  <publisher>IEEE</publisher>
  <pubdate>2019</pubdate>
  <volume>7</volume>
  <fpage>64722</fpage>
  <lpage>-64736</lpage>
</bibl>

<bibl id="B19">
  <title><p>Feature engineering for mid-price prediction with deep
  learning</p></title>
  <aug>
    <au><snm>Ntakaris</snm><fnm>A</fnm></au>
    <au><snm>Mirone</snm><fnm>G</fnm></au>
    <au><snm>Kanniainen</snm><fnm>J</fnm></au>
    <au><snm>Gabbouj</snm><fnm>M</fnm></au>
    <au><snm>Iosifidis</snm><fnm>A</fnm></au>
  </aug>
  <source>Ieee Access</source>
  <publisher>IEEE</publisher>
  <pubdate>2019</pubdate>
  <volume>7</volume>
  <fpage>82390</fpage>
  <lpage>-82412</lpage>
</bibl>

<bibl id="B20">
  <title><p>Mid-price prediction based on machine learning methods with
  technical and quantitative indicators</p></title>
  <aug>
    <au><snm>Ntakaris</snm><fnm>A</fnm></au>
    <au><snm>Kanniainen</snm><fnm>J</fnm></au>
    <au><snm>Gabbouj</snm><fnm>M</fnm></au>
    <au><snm>Iosifidis</snm><fnm>A</fnm></au>
  </aug>
  <source>Plos one</source>
  <publisher>Public Library of Science San Francisco, CA USA</publisher>
  <pubdate>2020</pubdate>
  <volume>15</volume>
  <issue>6</issue>
  <fpage>e0234107</fpage>
</bibl>

<bibl id="B21">
  <title><p>Functional data analysis</p></title>
  <aug>
    <au><snm>Ramsay</snm><fnm>JO</fnm></au>
  </aug>
  <source>Encyclopedia of Statistical Sciences</source>
  <publisher>Wiley Online Library</publisher>
  <pubdate>2004</pubdate>
  <volume>4</volume>
</bibl>

<bibl id="B22">
  <title><p>Applied functional data analysis: methods and case
  studies</p></title>
  <aug>
    <au><snm>Ramsay</snm><fnm>JO</fnm></au>
    <au><snm>Silverman</snm><fnm>BW</fnm></au>
  </aug>
  <publisher>Germany: Springer</publisher>
  <pubdate>2007</pubdate>
</bibl>

<bibl id="B23">
  <title><p>Introduction to functional data analysis</p></title>
  <aug>
    <au><snm>Kokoszka</snm><fnm>P</fnm></au>
    <au><snm>Reimherr</snm><fnm>M</fnm></au>
  </aug>
  <publisher>Boca Raton: CRC press</publisher>
  <pubdate>2017</pubdate>
</bibl>

<bibl id="B24">
  <title><p>Application of support vector machines in financial time series
  forecasting</p></title>
  <aug>
    <au><snm>Tay</snm><fnm>FE</fnm></au>
    <au><snm>Cao</snm><fnm>L</fnm></au>
  </aug>
  <source>omega</source>
  <publisher>Elsevier</publisher>
  <pubdate>2001</pubdate>
  <volume>29</volume>
  <issue>4</issue>
  <fpage>309</fpage>
  <lpage>-317</lpage>
</bibl>

<bibl id="B25">
  <title><p>Forecasting stock market movement direction with support vector
  machine</p></title>
  <aug>
    <au><snm>Huang</snm><fnm>W</fnm></au>
    <au><snm>Nakamori</snm><fnm>Y</fnm></au>
    <au><snm>Wang</snm><fnm>SY</fnm></au>
  </aug>
  <source>Computers \& operations research</source>
  <publisher>Elsevier</publisher>
  <pubdate>2005</pubdate>
  <volume>32</volume>
  <issue>10</issue>
  <fpage>2513</fpage>
  <lpage>-2522</lpage>
</bibl>

<bibl id="B26">
  <title><p>Kernel methods in finance</p></title>
  <aug>
    <au><snm>Chalup</snm><fnm>SK</fnm></au>
    <au><snm>Mitschele</snm><fnm>A</fnm></au>
  </aug>
  <source>Handbook on information technology in finance</source>
  <publisher>Germany: Springer</publisher>
  <pubdate>2008</pubdate>
  <fpage>655</fpage>
  <lpage>-687</lpage>
</bibl>

<bibl id="B27">
  <title><p>Regularization and variable selection via the elastic
  net</p></title>
  <aug>
    <au><snm>Zou</snm><fnm>H</fnm></au>
    <au><snm>Hastie</snm><fnm>T</fnm></au>
  </aug>
  <source>Journal of the royal statistical society: series B (statistical
  methodology)</source>
  <publisher>Wiley Online Library</publisher>
  <pubdate>2005</pubdate>
  <volume>67</volume>
  <issue>2</issue>
  <fpage>301</fpage>
  <lpage>-320</lpage>
</bibl>

<bibl id="B28">
  <title><p>Automation, speed, and stock market quality: The NYSE's
  hybrid</p></title>
  <aug>
    <au><snm>Hendershott</snm><fnm>T</fnm></au>
    <au><snm>Moulton</snm><fnm>PC</fnm></au>
  </aug>
  <source>Journal of Financial Markets</source>
  <publisher>Elsevier</publisher>
  <pubdate>2011</pubdate>
  <volume>14</volume>
  <issue>4</issue>
  <fpage>568</fpage>
  <lpage>-604</lpage>
</bibl>

<bibl id="B29">
  <title><p>Regularization Paths for Generalized Linear Models via Coordinate
  Descent</p></title>
  <aug>
    <au><snm>Friedman</snm><fnm>J</fnm></au>
    <au><snm>Hastie</snm><fnm>T</fnm></au>
    <au><snm>Tibshirani</snm><fnm>R</fnm></au>
  </aug>
  <source>Journal of Statistical Software</source>
  <pubdate>2010</pubdate>
  <volume>33</volume>
  <issue>1</issue>
  <fpage>1</fpage>
  <lpage>-22</lpage>
  <url>https://www.jstatsoft.org/v33/i01/</url>
</bibl>

<bibl id="B30">
  <title><p>Controlling the false discovery rate: a practical and powerful
  approach to multiple testing</p></title>
  <aug>
    <au><snm>Benjamini</snm><fnm>Y</fnm></au>
    <au><snm>Hochberg</snm><fnm>Y</fnm></au>
  </aug>
  <source>Journal of the Royal statistical society: series B
  (Methodological)</source>
  <publisher>Wiley Online Library</publisher>
  <pubdate>1995</pubdate>
  <volume>57</volume>
  <issue>1</issue>
  <fpage>289</fpage>
  <lpage>-300</lpage>
</bibl>

<bibl id="B31">
  <title><p>Federated Learning: Challenges, Methods, and Future
  Directions</p></title>
  <aug>
    <au><snm>Li</snm><fnm>T</fnm></au>
    <au><snm>Sahu</snm><fnm>AK</fnm></au>
    <au><snm>Talwalkar</snm><fnm>A</fnm></au>
    <au><snm>Smith</snm><fnm>V</fnm></au>
  </aug>
  <source>IEEE Signal Processing Magazine</source>
  <pubdate>2020</pubdate>
  <volume>37</volume>
  <issue>3</issue>
  <fpage>50</fpage>
  <lpage>60</lpage>
</bibl>

<bibl id="B32">
  <title><p>Advances and open problems in federated learning</p></title>
  <aug>
    <au><snm>Kairouz</snm><fnm>P</fnm></au>
    <au><snm>McMahan</snm><fnm>HB</fnm></au>
    <au><snm>Avent</snm><fnm>B</fnm></au>
    <au><snm>Bellet</snm><fnm>A</fnm></au>
    <au><snm>Bennis</snm><fnm>M</fnm></au>
    <au><snm>Bhagoji</snm><fnm>AN</fnm></au>
    <au><snm>Bonawitz</snm><fnm>K</fnm></au>
    <au><snm>Charles</snm><fnm>Z</fnm></au>
    <au><snm>Cormode</snm><fnm>G</fnm></au>
    <au><snm>Cummings</snm><fnm>R</fnm></au>
    <au><cnm>others</cnm></au>
  </aug>
  <source>Foundations and Trends{\textregistered} in Machine Learning</source>
  <publisher>Now Publishers, Inc.</publisher>
  <pubdate>2021</pubdate>
  <volume>14</volume>
  <issue>1--2</issue>
  <fpage>1</fpage>
  <lpage>-210</lpage>
</bibl>

</refgrp>
} 

\end{backmatter}
\end{document}